\documentclass[10pt,journal,compsoc]{IEEEtran}

\usepackage{soul, color}
\usepackage{url}

\usepackage[utf8]{inputenc}
\usepackage[small]{caption}
\usepackage{graphicx}
\usepackage{amsmath}
\usepackage{amsfonts}
\usepackage{booktabs}
\usepackage[flushleft]{threeparttable}
\usepackage{multirow}
\usepackage{tabularx}
\usepackage{xspace}
\usepackage{subcaption}
\usepackage{comment}
\usepackage{amsmath}
\usepackage{mathtools}
\usepackage{caption}
\usepackage{enumitem}
\usepackage{xr}

\usepackage{adjustbox}
\usepackage{varwidth}
\usepackage{xcolor}

\usepackage{amsthm}

\newcommand{\etal}{\mbox{\emph{et al.}}\xspace}

\newcommand{\method}{\mbox{$\mathop{\mathtt{HAM}}\limits$}\xspace}
\newcommand{\methodMax}{\mbox{$\mathop{\mathtt{HAM}_{x}}\limits$}\xspace}
\newcommand{\methodMean}{\mbox{$\mathop{\mathtt{HAM}_{m}}\limits$}\xspace}
\newcommand{\methodCor}{\mbox{$\mathop{\mathtt{HAM}^s}\limits$}\xspace}
\newcommand{\methodCorx}{\mbox{$\mathop{\mathtt{HAM^s_x}}\limits$}\xspace}
\newcommand{\methodCorm}{\mbox{$\mathop{\mathtt{HAM^s_m}}\limits$}\xspace}
\newcommand{\methodCormo}{\mbox{$\mathop{\mathtt{HAM^s_m\text{-}o}}\limits$}\xspace}
\newcommand{\methodCormu}{\mbox{$\mathop{\mathtt{HAM^s_m\text{-}u}}\limits$}\xspace}

\newcommand{\Caser}{\mbox{$\mathop{\mathtt{Caser}}\limits$}\xspace}
\newcommand{\SASRec}{\mbox{$\mathop{\mathtt{SASRec}}\limits$}\xspace}
\newcommand{\HGN}{\mbox{$\mathop{\mathtt{HGN}}\limits$}\xspace}
\newcommand{\GRURec}{\mbox{$\mathop{\mathtt{GRU4Rec}}\limits$}\xspace}
\newcommand{\GRURecP}{\mbox{$\mathop{\mathtt{GRU4Rec\text{++}}}\limits$}\xspace}
\newcommand{\NARM}{\mbox{$\mathop{\mathtt{NARM}}\limits$}\xspace}

\newcommand{\NextItRec}{\mbox{$\mathop{\mathtt{NextItRec}}\limits$}\xspace}

\newcommand{\CUT}{\mbox{$\mathop{\text{80-20-CUT}}\limits$}\xspace}
\newcommand{\CUS}{\mbox{$\mathop{\text{80-3-CUT}}\limits$}\xspace}
\newcommand{\LOS}{\mbox{$\mathop{\text{3-LOS}}\limits$}\xspace}

\newcommand{\se}{\mbox{$\mathop{\mathbf{h}}\limits$}\xspace}
\newcommand{\sse}{\mbox{$\mathop{\mathbf{o}}\limits$}\xspace}
\newcommand{\user}{\mbox{$\mathop{\mathbf{u}}\limits$}\xspace}
\newcommand{\she}{\mbox{$\mathop{\mathbf{s}}\limits$}\xspace}

\newcommand{\ESsub}{\mbox{$\mathop{\mathbf{s}^t_i}\limits$}\xspace}
\newcommand{\Esub}{\mbox{$\mathop{\mathbf{h}^t_i}\limits$}\xspace}
\newcommand{\Esubsub}{\mbox{$\mathop{\mathbf{o}^t_i}\limits$}\xspace}
\newcommand{\Elatest}{\mbox{$\mathop{a^i_t}\limits$}\xspace}

\newcommand{\E}{\mbox{$\mathop{\mathbf{v}}\limits$}\xspace}
\newcommand{\Q}{\mbox{$\mathop{\mathbf{w}}\limits$}\xspace}
\newcommand{\C}{\mbox{$\mathop{\mathbf{c}}\limits$}\xspace}
\newcommand{\EM}{\mbox{$\mathop{V}\limits$}\xspace}
\newcommand{\QM}{\mbox{$\mathop{W}\limits$}\xspace}
\newcommand{\U}{\mbox{$\mathop{U}\limits$}\xspace}

\newcommand{\rs}{\mbox{$\mathop{{r}}\limits$}\xspace}

\newcommand{\sequences}{\mbox{$\mathop{S}\limits$}\xspace}
\newcommand{\subsequences}{\mbox{$\mathop{S_i(t-l,l)}\limits$}\xspace}
\newcommand{\subsubsequences}{\mbox{$\mathop{S_i(t-1,1)}\limits$}\xspace}

\newcommand{\LL}{n_h}
\newcommand{\M}{n_l}
\newcommand{\T}{n_p}

\newcommand{\xia}[1]{\textcolor{red}{#1}}

\newcommand{\bo}[1]{\textcolor{magenta}{#1}}

\ifCLASSOPTIONcompsoc
  \usepackage[nocompress]{cite}
\else
  \usepackage{cite}
\fi

\begin{document}
\title{\method: Hybrid Associations Models\\for Sequential Recommendation}

\author{Bo~Peng,
        Zhiyun~Ren,
        Srinivasan~Parthasarathy,~\IEEEmembership{Member,~IEEE,}
        and~Xia~Ning$^*$,~\IEEEmembership{Member,~IEEE}
\IEEEcompsocitemizethanks{
\IEEEcompsocthanksitem Bo Peng 
is with the Department
of Computer Science and Engineering, The Ohio State University, Columbus,
OH, 43210.\protect\\
E-mail: peng.707@buckeyemail.osu.edu 
\IEEEcompsocthanksitem Srinivasan Parthasarathy and Xia Ning are with the Department of Biomedical Informatics, 
and the Department of Computer Science and Engineering, 
The Ohio State University, Columbus, OH, 43210.\protect\\
E-mail:  srini@cse.ohio-state.edu, ning.104@osu.edu
\IEEEcompsocthanksitem Zhiyun Ren is with the Department of Biomedical Informatics, 
The Ohio State University, Columbus, OH, 43210.\protect\\
E-mail: ren.685@osu.edu
\IEEEcompsocthanksitem $^*$Corresponding author\protect\\
\IEEEcompsocthanksitem This work has been submitted to the IEEE for possible publication.
Copyright may be transferred without notice, after which this version may no longer be accessible.
}
\thanks{Manuscript received April 19, 2005; revised August 26, 2015.}}

\markboth{Journal of \LaTeX\ Class Files,~Vol.~14, No.~8, August~2015}%
{Shell \MakeLowercase{\textit{et al.}}: Bare Demo of IEEEtran.cls for Computer Society Journals}
%


\graphicspath{{./plots/data_distribution/}{./plots/CDs/}{./plots/Comics/}{./plots/ml1m/}{./plots/ml20m/}}

\IEEEtitleabstractindextext{%
\begin{abstract}
Sequential recommendation aims to identify and recommend 
the next few items for a user that the user is most likely to purchase/review, given the user's 
purchase/rating trajectories. It becomes an effective tool to  
help users select favorite items from a variety of options.  
In this manuscript, we developed hybrid associations models (\method) 
to generate sequential recommendations using three factors: 
1) users’ long-term preferences,
2) sequential, high-order and low-order association patterns in the users' most recent purchases/ratings, and
3) synergies among those items.
\method uses simplistic pooling to represent a set of items in the associations, 
and element-wise product to represent item synergies of arbitrary orders. 
%
We compared \method  models
with the most recent, state-of-the-art methods on six public benchmark datasets in three
 different experimental settings. 
Our experimental results demonstrate that  
{\method} models significantly outperform the state of the art in all the experimental settings, 
with an improvement as much as 46.6\%.
In addition, our run-time performance comparison in testing demonstrates 
that \method models are much more efficient than the state-of-the-art methods, and are able to 
achieve significant speedup as much as 139.7 folds.
%
%
%
%

\end{abstract}

}

\maketitle

\IEEEdisplaynontitleabstractindextext
\IEEEpeerreviewmaketitle

\IEEEraisesectionheading{\section{Introduction}\label{sec:introduction}}
\IEEEPARstart{S}{equential} recommendation aims to identify and recommend 
the next few items for a user that the user is most likely to purchase/review, given the user's 
purchase/rating trajectories (e.g., sequences of purchased items, and time stamps of the purchases if available). 
It becomes an effective tool to  
help users select favorite items from a variety of options.   
A key challenge in sequential recommendation is to identify, learn or represent the patterns and dynamics 
in users' purchase/rating sequences that are most pertinent to inform their future 
interactions with other items, and also to capture the relations between such patterns and 
future interactions. 
With the prosperity of deep learning, many deep models, particularly based on 
recurrent neural networks~\cite{hidasi2015session,hidasi2018recurrent} and with 
attention mechanisms, have been developed for sequential recommendation purposes.  
These methods typically model users' sequential behaviors and their 
long-term/short-term preferences, and have significantly improved
recommendation performance. 
%

However, given the notoriously sparse nature of data in recommendation problems, 
a question on these deep learning methods, particularly those with attention mechanisms, is that whether 
the sparse recommendation data are sufficient to 
enable well-learned attention weights or any hypothesized recurrent 
patterns with many parameters, 
and whether these learned weights and patterns can
play effective roles in identifying important information  
leading to accurate recommendations. 
%
%
%
%
Recent studies~\cite{dacrema2019we,ludewig2020} bring such concerns on recommendation algorithms in general, 
demonstrating that complicated deep recommendation methods may not always outperform simple ones. 
We have similar conclusions based on our study on the 
attention weights of a state-of-the-art method~\cite{ma2019hierarchical} in Section~\ref{sec:discussion:weight}: 
the learned attention weights are not always meaningful.

To better learn from sparse, historical purchase/rating sequences 
and generate more accurate sequential recommendations, we ask the questions whether simpler models are 
actually more effective and more efficient.
Particularly, we wonder whether simplistic pooling instead of attention mechanisms, and more explicit and intuitive patterns 
instead of highly parameterized recurrence are superior on sparse, sequential 
recommendation data. 
Motivated by the above questions, 
we propose a set of new, hybrid associations models, denoted as \method. 
%
\method models explicitly use the following three intuitive factors to generate the sequential recommendations 
via a simplistic linear scoring function:  
1) users' general/long-term preferences, 
2) sequential, association patterns in the users' most recent purchases/ratings, and 
3) synergies among the most recent, purchased/rated items.
It has been shown that  
the users' general preferences persist in a relatively long period~\cite{rendle2010factorizing}, and 
play an important role to determine in their purchases/ratings~\cite{he2016fusing}. %
For example, some users may prefer warm colors, while others may 
prefer cool colors; such preference would effect their purchases during a long period. 
Previous study~\cite{tang2018personalized} has also shown  
the existence of sequential associations in recommendation data (e.g., sequential association rules). Thus, among all the items,  
the probabilities of being purchased next given previous purchases will not be equal for each user. 
Moreover, 
the synergies among items, also called union-level effects in literature~\cite{tang2018personalized}, could
enable helpful, additional information for the recommendations that each individual items could not on their own.
%

In \method models, 
we learn the users' general preferences by leveraging their all historical purchases/ratings
and represent the preferences in user embeddings. Thus, the user embeddings will encode the 
user preferences that are relatively consistent across different items. 
The item association patterns 
used in \method include both high-order associations (i.e., many items together induce the next, a few items)
and low-order associations (i.e., only a few items together induce the next, a few items).
We use the pooling mechanism (i.e., mean pooling and max pooling)
to model the high/low-order item associations. 
The \method model with item associations and max/mean pooling is denoted as \methodMax/\methodMean. 
%
In addition, we explicitly model the synergies among item pairs
using Hadamard product from their embeddings and 
extend that to model the synergies among arbitrary numbers 
of items in a recursive way.
We use the latent cross~\cite{beutel2018latent} technique to effectively
combine the item synergies and item associations. 
%
%
The \method model with both item associations and synergies among the associations is denoted as \methodCor. 
The
recommendations on the next items are generated based on their recommendation scores aggregated from users' general 
preferences, item association patterns (and item synergies). 

We conducted a comprehensive set of experiments and compare the \method models 
with 3 the most recent, state-of-the-art baseline methods on 6 public benchmark datasets in 3 different experimental settings.  
%
%
Our experimental results demonstrate that  
{\method} models significantly outperform the state-of-the-art methods in all the experimental settings, 
with the best improvement 13.3\% over 
the state-of-the-art method {\HGN}.  
%
In addition, we conducted an ablation study to verify the effects of low-order associations 
and users' general preferences.
The results demonstrate the importance of explicitly modeling both of them. 
Moreover, we compared the {run-time performance in testing} of \method models and baseline methods. 
The results show that \method  models are much more efficient than deep methods 
with a speedup as much as 139.7 folds, and an average 2.2-fold speedup compared to the 
state of the art. 
%
%
%
%


%

We also conducted a thorough study on all the baseline methods with exhaustive parameter tuning. 
%
We studied the attention weights learned in benchmark datasets by one of the state-of-the-art methods, 
and discussed the potential issues of using attention mechanisms in 
sequential recommendation. 
We also studied various experimental settings for
evaluating the performance of sequential recommendation and 
discussed the potential issues. 

Our major contributions are summarized as follows:
\begin{itemize}[noitemsep,nolistsep,leftmargin=*]
\item We proposed a novel sequential recommendation method \method. \method  
models three intuitive factors explicitly 
1) users' long-term preference, 
2) sequential, association patterns among items 
and 3) synergies among items. 
To the best of our knowledge, \method is the first method that explicitly models both 
high-order and low-order sequential associations among items, and their synergies  
(Section~\ref{sec:methods}).
\item \method uses simplistic pooling instead of learned attentions to efficiently 
represent a set of items 
(Section~\ref{sec:methods}).
%
\item
\method significantly outperforms the state-of-the-art methods. 
The experimental results over 6 benchmark datasets demonstrate that \method 
achieves significant improvement in both 
recommendation performance and time efficiency compared to 
the state-of-the-art methods in various experimental settings 
(Section~\ref{sec:results}).
\item We studied various experimental settings in which sequential recommendation performance is 
evaluated, and discussed the potential issues in the most widely used experimental setting in literature 
(Section~\ref{sec:discussion}).
\item 
We conducted a study on the state-of-the-art methods more thoroughly 
than reported in their respective papers, 
and discussed potential issues 
(Section~\ref{sec:discussion}).
\item We investigated the attention weights learned in benchmark datasets and studied the potential 
scenarios in which pooling could outperform attention mechanisms in sequential recommendation 
(Section~\ref{sec:discussion}).
\item For reproducibility and fair comparison, we published our code and data 
at~\url{https://github.com/BoPeng112/HAM} and reported all 
the parameters used in the study (Appendix).
\end{itemize}
%

\section{Literature Review}
\label{sec:review}

Numerous sequential recommendation methods have been developed, particularly
using  
Markov Chains (MCs), Recurrent Neural Networks (RNNs),
Convolutional Neural Networks (CNNs), attention and gating mechanisms, etc. 
Specifically, MCs-based methods, such as factorized personalized Markov chains method 
(FPMC)~\cite{rendle2010factorizing}, use  MCs to capture  pairwise 
item-item transition relations to 
recommend the next item for each user.
Based on FPMC, He \etal~{\cite{he2016fusing}} used
 high-order MCs to capture the impact from all historical purchases/ratings
on the next item. 
Recently, many RNN-based methods~\cite{hidasi2015session,hidasi2018recurrent,quadrana2017personalizing} have been developed to model the sequential 
patterns in users' purchase/rating sequences.
For instance, Hidasi \etal~\cite{hidasi2015session} developed a gated recurrent units (GRUs) 
based recommendation model (\GRURec), 
which uses GRUs to capture users' short-term preferences. 
Hidasi \etal~\cite{hidasi2018recurrent} further improved \GRURec to \GRURecP
by introducing a novel ranking loss to mitigate the 
gradient vanish problem in GRUs.
In addition, Quadrana \etal~\cite{quadrana2017personalizing} developed a hierarchical RNN-based 
recommendation model to capture 
the users' short-term preferences in the current session and 
the evolvement of users' preferences across different sessions.

Besides RNNs, attention mechanisms are also adapted to 
model the sequential patterns among sequences.
Li \etal~\cite{li2017neural} developed a neural attentive 
recommendation model (\NARM), 
which extends \GRURec by introducing the attention mechanism to 
improve the GRUs' ability on modeling long-term dependence.
Liu \etal~\cite{liu2018stamp} developed an attention-based 
recommendation mode that solely relies on attention mechanisms 
to model the sequential patterns among sequences.
Kang \etal~\cite{kang2018self} developed a self-attention-based sequential 
model (\SASRec) to capture a few, most informative items in users' purchase/rating   
sequences to generate recommendations. 
Recent work also adapts CNNs for sequential recommendation. 
For example, Tang \etal~{\cite{tang2018personalized}} developed 
a convolutional sequence embedding recommendation model (\Caser),  
which uses multiple convolutional filters on the most recent 
purchases/ratings to extract union-level features from a set of recent items.
Yuan \etal~\cite{yuan2019simple} developed a CNN-based generative model \NextItRec to 
improve \Caser on learning long-term dependencies in the item sequences.
%
Ma \etal~\cite{ma2019hierarchical} developed a hierarchical gating network (\HGN),   
which captures item-item transition relations and user long-term preferences, and
uses gating mechanisms to identify important items and their latent features
from users' historical purchases/ratings.  
\HGN has been demonstrated as the state of the art, and 
outperforms an extensive set of existing methods including \GRURec~\cite{hidasi2015session}, 
\GRURecP~\cite{hidasi2018recurrent} and \NextItRec~\cite{yuan2019simple}.

\section{Definitions and Notations}
\label{sec:notations}


Table~\ref{tbl:notations} presents the key notations using in this manuscript. 
In this manuscript, the historical purchases or ratings of user $i$ in chronological order are represented as a sequence 
$\sequences_i$=$\{s_i(1), s_i(2), \cdots\}$, 
where $s_i(t)$ is the $t$-th purchased/rated item.
%
A length-$l$ subsequence of $S_i$ starting at the $t$-th purchase/rating is denoted as $S_i(t, l)$, 
that is, $S_i(t, l) = \{s_i(t), s_i(t$$+$$1), \cdots, s_i(t$$+$$l$$-$$1)\}$. When no ambiguity arises, we will eliminate 
$i$ in $S_i$/$S_i(t, l)$. 
We use upper-case letters to denote matrices, 
lower-case and bold letters to denote row vectors, 
and lower-case non-bold letters to represent scalars.
In this manuscript, we tackle the sequential recommendation problem defined as follows: 

\vspace{0.5em}
\noindent
\textit{\textbf{Sequential Recommendation Problem}: 
for a user $i$, given $\sequences_i$, recommend a few items she/he will purchase/rate next. }


\section{Methods}
\label{sec:methods}

\begin{figure}[!t]
    \centering
    \includegraphics[width=1.0\linewidth]{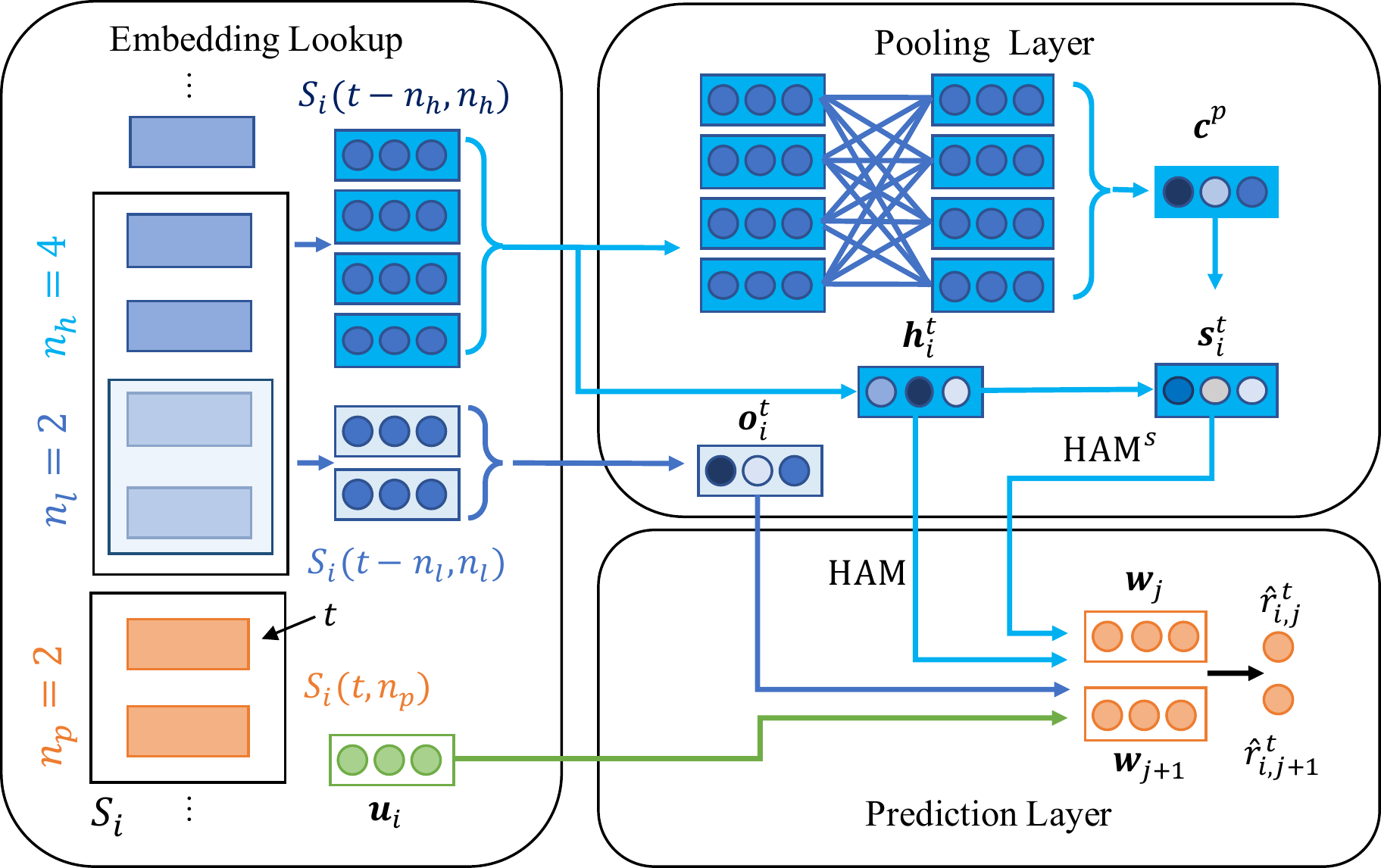}
    \captionof{figure}{\method Model Architecture}
        \label{fig:method}
        \vspace{-10pt}
\end{figure}

Fig.~\ref{fig:method} presents the \method architecture. 
%
%
{\method} generates recommendations for the next items for each user using three factors: 
1) the user's general/long-term preferences (Section~\ref{sec:method:general}), 
2) the sequential, association patterns in the 
user's most recent purchases/ratings (Section~\ref{sec:method:sequential}), and
3) synergies among items that users purchased/rated most recently 
(Section~\ref{sec:method:sequential:highLevel}).
These three factors will be used to calculate a recommendation
score for each item candidate (Section~\ref{sec:method:prediction}) 
in order to prioritize and recommend the next items.
%


Before we discuss \method, we first discuss the heterogeneous item embeddings here that \method will be using. 
It has been shown~\cite{rendle2010factorizing,kang2018self,kabbur2013fism} that the item transitions may 
be asymmetric: 
item $j$ might be frequently purchased/rated after item $k$, but not vise versa.
To model such asymmetry in \method, 
we follow the ideas of heterogeneous item embeddings~\cite{kang2018self}
and learn two item embedding matrices, denoted as $\EM \in \mathbb{R}^{n \times d}$ 
and $\QM \in \mathbb{R}^{n \times d}$, respectively.
%
Both \EM and \QM are lookup tables, in which the $j$-th rows, denoted as $\E_j$ and $\Q_j$,
respectively, represent item $j$.
In \method, 
if item $j$ is used to recommend the next items, 
it is represented by $\E_j$.
If item $j$ is a candidate to be recommended,
it is represented by $\Q_j$.

\begin{table}[!t]
%
  \centering
  \caption{Notations}
  \label{tbl:notations}
  \begin{threeparttable}
      \begin{tabular}{
        @{\hspace{0pt}}l@{\hspace{1pt}}
        @{\hspace{0pt}}p{0.85\linewidth}@{\hspace{0pt}}
        }
        \toprule
        notations & \hspace{5pt}meanings \\
        \midrule
        $m$/$n$, $d$ &  number of users /items, embedding dimension\\
        $\sequences_i$ & purchase/rating sequence of user $i$ \\
        $\sequences_i(t, l)$ & a length-$l$ subsequence of $\sequences_i$ starting
         from the $t$-th purchase/rating\\
        \U/\EM &  user/item embedding matrix \\
        \QM &  candidate item embedding matrix \\
        $\rs_{ij}$ & the recommendation score of user $i$ 
                     on item $j$ \\
        $n_h$/$n_l$ & the number of items in high-/low-order  associations \\
        {$n_p$}  & the number of items to calculate recommendation errors during training \\
       \bottomrule
      \end{tabular}
  \end{threeparttable}
  \vspace{-10pt}
\end{table}


\subsection{Modeling Users' General Preferences}
\label{sec:method:general}

It has been shown that 
the users' general preferences persist in a relatively long period~\cite{he2018adversarial},
and
play an important role in users' purchases/ratings~\cite{he2016fusing}.
For example, some users may prefer items of low price, while others may prefer luxurious items
that could be more expensive.
%
Therefore, in \method, 
we learn users' general preferences using an embedding matrix $\U \in \mathbb{R}^{m\times d}$ 
(optimization details in Section~\ref{sec:method:obj}). 
\U serves as a lookup table, in which the $i$-th row, denoted as $\user_i$, 
represents the general preferences of user $i$.
%
Importantly, $\user_i$ stays the same for user $i$'s recommendations, and is learned from all the 
items purchased/rated by the user. 
%
%

\subsection{Modeling Sequential Hybrid Associations}
\label{sec:method:sequential}

Previous study~\cite{tang2018personalized} has shown  
the existence of sequential associations in data of recommendation problems.  
That is, 
the probabilities of being purchased next given previous purchases will not be equal for all the items for each user. 
For example, after watching the movie Avengers: Infinity War, 
users may be more likely to watch the movie Avengers: Endgame, which is the sequel of the previous movie, 
than the movie Roman Holiday, which is of a different genre and story.
Therefore, using the sequential, association patterns, we can better identify the next most possible items given 
immediately previous purchases. 

We denote the sequential association at time $t$ from 
its previous $\LL$ purchases/ratings to the next $\T$ subsequent purchases/ratings
as 
$S(t$$-$$\LL, \LL)$$\rightarrow$$S(t, \T)$, 
and the number of the involved $\LL$$+$$\T$ items as the  order of the association. 
%
%
It has also been shown~\cite{tang2018personalized} that the sequential associations among items 
in benchmark recommendation datasets (used in experiments as in Section~\ref{sec:experiments:datasets}) have different orders.
For example, about 50\% significant associations have $\LL$$=$$2$ and $\T$$\leq$$2$
(i.e., previous 2 purchases/ratings have immediate effects on the next 1 or 2 purchases/ratings),
and about 15\% significant associations have $\LL$$=$$4$ and $\T$$\leq$$2$. 
%
%
%

We propose to explicitly model the item associations of different orders in \method such that 
the aggregated information from the previous different numbers of purchases/ratings will contribute to the
recommendation scores of all the subsequent item candidates. 
Particularly, we include a high-order 
association $S(t$$-$$\LL, \LL)$$\rightarrow$$S(t, \T)$ and a low-order association 
$S(t$$-$$\M, \M)$$\rightarrow$$S(t, \T)$, where $\M$$<$$\LL$ and $S(t$$-$$\M, \M)$$\subset$$S(t$$-$$\LL, \LL)$, to recommend 
the next $\T$ items. 
During training, we split the historical purchase/rating sequence of each user into 
multiple subsequences (training instances) of length $\LL$+$\T$ (these subsequences will overlap through a sliding window, 
Section~\ref{sec:experiments:setting}), 
and learn the individual item embeddings, and thus the embeddings of $\LL$/$\M$ successive items, from 
all the length-($\LL$+$\T$) subsequences.
Note that \method can be a general framework, in which arbitrary numbers of 
various-order associations can be incorporated. 
We use high-order associations and low-order associations in \method because 
intuitively, the most recent purchased/rated items are more informative 
than earlier items in indicating the next items that users may purchase/rate.
Thus, we emphasize the effect of the most recent purchased/rated items in 
$S(t$$-$$\M, \M)$ by explicitly modeling the low-order associations, and also use $S(t$$-$$\LL, \LL)$
to incorporate a slightly longer purchase history and its trend. 

\subsubsection{Modeling Item Associations: \methodMax and \methodMean}
\label{sec:method:sequential:pooling}
In order to represent the information from the previous $\LL$/$\M$ purchased/rated items as a whole, 
we use pooling mechanisms, and learn one embedding $\se$/$\sse$ for the $\LL$/$\M$ successive
purchases/ratings using 1) max pooling and 2) mean pooling, respectively,
from individual item embeddings of the $\LL$/$\M$ purchased/rated items, that is, 
\begin{eqnarray}
\label{eqn:maxh}
\begin{aligned}
& \text{for $S_i(t$$-$$\LL,\LL)$: } && \Esub = \underset{{j \in S(t-\LL,\LL)}}{\text{max/mean}}(\E_j), \\
& \text{for $S_i(t$$-$$\M,\M)$:   } && \Esubsub = \underset{{j \in S(t-\M,\M)}}{\text{max/mean}}(\E_j), 
\end{aligned}
\end{eqnarray}
%
%
%
%
where $\text{max/mean}$ indicates either max pooling or mean pooling. 
\method with high-order and low-order associations calculated from the above 
max and mean pooling is denoted as \methodMax and \methodMean, respectively.

Given the sparse nature of the data in recommendation problems, it might be difficult to learn and differentiate
the contributions of different items.
The mean pooling is a simplistic solution to average the 
effects from each individual item. 
The max pooling is based on the assumption that the purchased/rated items contribute in various dimensions
to determine the next $\T$ purchases/ratings.

\subsubsection{\mbox{Modeling High-Order Item Synergies: \methodCorx and \methodCorm}}
\label{sec:method:sequential:highLevel}


The pooling in Section~\ref{sec:method:sequential:pooling} learns the aggregated effects of 
all the items together in $S_i(t$$-$$\LL,\LL)$ and in $S_i(t$$-$$\M,\M)$ on recommending next items. 
Literature suggests 
that there could be synergies among certain items~\cite{tang2018personalized}
that could impact the next items differently
(so called union-level effects).
For example, if the user purchased candles and wines together, 
it is likely that she/he is preparing for a candlelight dinner, and thus we may recommend steaks to her/him 
for the dinner. However, if we consider candles or wines independently, better recommendations could be 
birthday cakes or beer, based on item associations.
Thus, the synergies among items in $S_i(t$$-$$\LL,\LL)$ 
may provide additional information that is not possible from individual items alone. 
 Motivated by this observation, we further 
incorporate the synergies among the items in \method. 

In \method, we model the synergies among  item $j$ and item $k$ in $S(t$$-$$\LL,\LL)$
using Hadamard product (element-wise product) of their embeddings as follows: 
\begin{equation}
    \label{eqn:cor}
    \C_{jk}^{(2)} = \E_j \circ \E_k,
\end{equation}
where $\C_{jk}^{(2)}$ is the synergy vector between the items $j$ and $k$; 
$\E_j$ and $\E_k$ are the embeddings of item $j$ and $k$, 
respectively; $\circ$ is the Hadamard product operator (element-wise product).
The intuition is that if two items both have high values on some latent embedding dimensions, 
they may have similar properties corresponding to those dimensions. The element-wise product 
will strengthen those properties, whereas mean pooling otherwise could smooth them out.

Given the pairwise synergies calculated as in Equation~\ref{eqn:cor}, 
we aggregate all the synergies between item $j$ and all the other items in $S(t-\LL, \LL)$ as follows:
%
\begin{equation}
    \label{eqn:cor2}
     \C^{(2)}_j  =  \sum_{k \in S(t-\LL, \LL), k\neq j} \C_{jk}^{(2)},  
\end{equation}
and aggregate all the synergies in $S_i(t$$-$$\LL,\LL)$ as follows: 
\begin{equation}
    \label{eqn:cor3}
        \C^{(2)}     =  \underset{j \in S(t-\LL, \LL)}{\text{mean}} (\C_{j}^{(2)}).
\end{equation}
%
%
%
The above synergies only consider pairs of items (i.e., order-2, Equation~\ref{eqn:cor}). It is possible that 
synergies exist among multiple items (i.e., order-$p$, $p$$>$$2$). 
In order to model synergies among arbitrary numbers of items, 
inspired by the idea in Min~\etal~\cite{min2014interpretable}, we
extend the order-$2$ synergies to order-$p$ synergies in a recursive way as follows:
\begin{equation}
    \label{eqn:corn}
    \C^{(p)} = \underset{j \in S(t-\LL, \LL)}{\text{mean}} (\sum_{k \in S(t-\LL, \LL), k\neq j} \C_j^{(p-1)} \circ \E_k),
    \vspace{-2pt}
\end{equation}
where $p\le n_h$, $\C_j^{(1)}$$=$$\E_j$ and $\C^{(p)}$ represents all the synergies among $p$ items.
That is, the 
order-$p$ synergies are constructed from order-$(p\text{-}1)$ synergies and a single item. 

Using $\C^{(p)}$ and following the idea of latent cross~\cite{beutel2018latent}, we combine
the information of individual items and synergies among items to represent high-order associations in 
\mbox{$S_i(t-\LL, \LL)$} for user $i$ as follows:
\vspace{-2pt}
\begin{equation}
    \label{eqn:cross}
    \ESsub = \Esub + \sum_{k=2}^p \C^{(k)} \circ \Esub, 
\end{equation}
where $\Esub$ is the embedding of $\LL$ items 
calculated from item pooling (Equation~\ref{eqn:maxh}); the latent cross term $\C^{(k)} \circ \Esub$
utilizes the information from item synergies to strengthen important latent features in $\Esub$; 
a new item association embedding $\ESsub$ is thus calculated from the latent cross and the high-order 
item association embedding. 
\method with high-order synergies is denoted as \methodCor. 
\method with high-order synergies and item associations calculated from   
max and mean pooling (Equation~\ref{eqn:maxh}) is denoted as \methodCorx and \methodCorm,
respectively.

Please note that we don't model the synergies among items in $S_i(t$$-$$\M,\M)$ (i.e., among low-order associations)
since we empirically found that \method generally achieves the best performance 
when $\M$ is small (i.e., 1 or 2) (will be discussed later in Section~\ref{sec:results:parameter}). 
Thus, we assume the synergies among very a few items are not significant enough to substantially improve
recommendation performance.
Actually, we empirically observed that it is the case in our experiments. 
%
Please also note that multiple combinations of pooling methods such as weighted sum
or max pooling could be used in Equation~\ref{eqn:cor2} and Equation~\ref{eqn:cor3}.
We tried weighted sum and max pooling in our experiments but 
empirically found that sum in Equation~\ref{eqn:cor2} and mean pooling in Equation~\ref{eqn:cor3}
work best.
The possible reason could be that 
sum will aggregate item synergies but not smooth them out.

\subsection{Recommendation Scores}
\label{sec:method:prediction}

The recommendation scores are calculated from the user embedding $\user$ (i.e., user's general preference, 
Section~\ref{sec:method:general}), 
the embedding of previous $\LL$ purchases/ratings (i.e., high-order association) $\se$/$\she$ 
(Equation~\ref{eqn:maxh} or \ref{eqn:cross})
and the embedding of previous $\M$ purchases/ratings (i.e., low-order association) \sse (Equation~\ref{eqn:maxh}).
For user $i$, given the subsequence $S_i(t$$-$$\LL,\LL)$, 
the recommendation score of user $i$ on 
item candidate $j$, denoted as $\rs^t_{ij}$, is calculated as follows:
\begin{equation}
    \label{eqn:prediction}
    \text{in \method: }\rs^t_{ij} = ~~\underbrace{\user_i \Q_j^\top}_{\mathclap{\hspace{-60pt}\text{user's general preferences}}} 
    \quad~~+\quad
    \underbrace{\Esub \Q_j^\top}_{{\hspace{-30pt}\text{high-order association}}} 
    + 
    \underbrace{\Esubsub \Q_j^\top,}_{\hspace{-5pt}\text{low-order association}} 
   \vspace{-6pt}
\end{equation}
%
%
\begin{equation}
\vspace{-3pt}
    \label{eqn:predictionS}
    \text{in \methodCor: }\rs^t_{ij} = ~~\underbrace{\user_i \Q_j^\top}_{\mathclap{\hspace{-70pt}\text{user's general preferences}}} 
    \quad~~+\quad
    \underbrace{\ESsub \Q_j^\top}_{{\hspace{-30pt}\substack{\text{high-order association}\\\text{with synergies}}}} 
    + 
    \underbrace{\Esubsub \Q_j^\top,}_{\hspace{-5pt}\text{low-order association}} 
\end{equation}
%
%
where $\Q_j$ 
is the embedding of candidate item $j$, 
\Esub/\ESsub is the embedding for $S_i(t$$-$$\LL,\LL)$ without/with synergies,  
and \Esubsub is the embedding for $S_i(t$$-$$\M,\M)$; 
$\user_i \Q_j^\top$ measures how item candidate $j$ matches 
user $i$'s general preferences; 
$\Esub \Q_j^\top$/$\ESsub \Q_j^\top$ measures how $S_i(t$$-$$\LL,\LL)$ induces item candidate $j$, 
and $\Esubsub \Q_j^\top$ measures how $S_i(t$$-$$\M,\M)$ induces item candidate $j$.
For each user, we recommend the items of {top-$k$} largest recommendation scores.
Note that we do not explicitly weight the three factors, as 
their weights can be learned as part of the user/sequence/item embeddings. 

\subsection{\method Optimization}
\label{sec:method:obj}

We adapt the Bayesian personalized ranking objective~\cite{rendle2012bpr}
and minimize the loss that occurs when the truly purchased/rated items are ranked below those not purchased/rated
based on the recommendation scores. 
The objective function for \method models is as follows: 
%
%
%
\begin{equation}
    \label{eqn:obj}
    \min\limits_{\boldsymbol{\Theta}} \sum^m_{i=1} 
        \sum_{\scriptsize{{\sequences_i(t,\T)  \subset \sequences_i}}}~~~~~~~~
    \sum_{\mathclap{~~~~~~~~~~~~~{\scriptsize{j \in \sequences_i(t, \T)} \scriptsize{k \notin \sequences_i(t, \T)}}}}
    -\log\sigma(\rs_{ij} - \rs_{ik}) + \lambda(\|\boldsymbol{\Theta}\|^2),
\end{equation}
where $\boldsymbol{\Theta} = \{\U, \EM, \QM\}$ is the set of the parameters; 
$\sigma$ is the sigmoid function;  
$\sequences_i(t, \T)$ is a  sequence of $\T$ items in $\sequences_i$;  
$j$ denotes an item in $\sequences_i(t,\T)$; 
%
%
and $k$ denotes an item not in $\sequences_i(t,\T)$. 
Given the huge number of items not in $\sequences_i(t,\T)$,
following the ideas in literature~\cite{tang2018personalized,ma2019hierarchical}, 
we randomly sample a non-purchased/rated item $k$ for each purchased/rated item $j$.
%
%
%
Please note that, in \method, the recommendation scores of purchased/rated item 
are not necessarily close to their ground-truth ratings, 
as long as the scores of purchased/rated items are higher 
than scores of those not purchased/rated.
%
Also, note that each of the items in $\sequences_i(t,\T)$ will be recommended and evaluated 
independently,  
following the literature~\cite{ma2019hierarchical}. 
%
We use Adam optimizer~\cite{kingma2014adam} to
optimize the objective function.
%

\section{Materials}
\label{sec:materials}

\subsection{Baseline Methods}
\label{sec:experiments:baseline}

We compare \method with three state-of-the-art methods:

\begin{itemize}[noitemsep,nolistsep,leftmargin=*]
\item 
\textbf{\Caser}\cite{tang2018personalized}:  
This method uses multiple convolutional filters on the most recent purchases/ratings  
of a user to extract the user's sequential features and items' group features.  
These two features and the users' long-term preferences are used to calculate item 
recommendation scores. 
\item 
\textbf{\SASRec}~\cite{kang2018self}:
This method uses self-attention mechanisms to capture the most informative 
items in users' purchase/rating sequences to recommend the next item. 
%
%
\item 
\textbf{\HGN}~\cite{ma2019hierarchical}: 
This method uses gating mechanisms to identify important items and their latent features 
from users' historical purchases/ratings to recommend next items.  
\end{itemize}

%
%

Note that 
\HGN has been compared with a comprehensive set 
of other methods including \GRURec~\cite{hidasi2015session}, 
\GRURecP~\cite{hidasi2018recurrent}, \NextItRec~\cite{yuan2019simple}
and has been demonstrated outperforming those methods. 
Thus, we compare \method 
with \HGN instead of the methods that \HGN outperforms.
For all the baseline methods, we use the implementation 
provided by the authors in github 
(Section~\ref{app:sec:para} in the Appendix).

\subsection{Datasets}
\label{sec:experiments:datasets}

We evaluate the methods on 6 public benchmark datasets:  
Amazon-Books (Books) and Amazon-CDs (CDs)~\cite{he2016ups}, 
Goodreads-Children (Children) and Goodreads-Comics (Comics)~\cite{wan2018item}, and 
MovieLens-1M (ML-1M) and MovieLens-20M (ML-20M)~\cite{harper2016movielens}.   
The Books and CDs datasets are from Amazon reviews~\cite{amazon}, 
which contain users' 1-5 star ratings and reviews on books and CDs, respectively. 
%
%
The Children and Comics datasets are from goodreads website~\cite{goodreads}.
These two datasets have users' implicit feedback (i.e., if a user has read the book or not), 
explicit feedback (i.e., ratings) and reviews on children and comics books.
The ML-1M and ML-20M datasets are from the MovieLens website~\cite{movielens}
with user-movie ratings. 
%
%
Following the data preprocessing protocol in \HGN~\cite{ma2019hierarchical}, 
among the 6 datasets, we only kept the users with at least 10 ratings, and items with at least 5 ratings. 
%
We converted the rating values into binary values by setting rating 4 and 5 to value 1, and the 
lower ratings to value 0.
Table~\ref{tbl:dataset} presents the statistics of the 6 datasets. 

\begin{table}[t!]
\small
  \caption{Dataset Statistics}
  \centering
  \label{tbl:dataset}
  \begin{threeparttable}
     \begin{footnotesize}
      \begin{tabular}{
	@{\hspace{0pt}}l@{\hspace{10pt}}
	@{\hspace{10pt}}r@{\hspace{10pt}}          
	@{\hspace{10pt}}r@{\hspace{10pt}}
	@{\hspace{10pt}}r@{\hspace{6pt}}
	@{\hspace{2pt}}r@{\hspace{10pt}}
        @{\hspace{10pt}}r@{\hspace{0pt}}
	}
        \toprule
        dataset & \#users   & \#items &  \#intrns & {\#intrns/u} & \#u/i\\
        \midrule
        CDs       & 17,052  & 35,118  & 472,265  & 27.7 & 13.4 \\
        Books    & 52,406   & 41,264 & 1,856,747 & 35.4 & 45.0 \\
        Children & 48,296  & 32,871 & 2,784,423  & 57.6 & 84.7 \\
        Comics  & 34,445  &  33,121 &  2,411,314 & 70.0 & 72.8 \\
        ML-20M & 129,780 & 13,663 & 9,926,480    & 76.5 & 726.5\\
        ML-1M   &  5,950    & 3,125   & 573,726  & 96.4 & 183.6\\
        \bottomrule
      \end{tabular}
      \end{footnotesize}
      \begin{tablenotes}[normal,flushleft]
      \begin{scriptsize}
      \item The columns ``\#users"/``\#items"/``\#intrns" represent the number of 
      users/items/user-item interactions, respectively. 
      The column of ``\#intrns/u" represents the average 
      number of interactions (average length of purchase/rating sequence) of each user. 
      The column of ``\#u/i" represents the average number of purchases/ratings on each item.  
      \par
      \end{scriptsize}
      \end{tablenotes}
  \end{threeparttable}
  \vspace{-5pt}
\end{table}


As in Table~\ref{tbl:dataset}, CDs is the most sparse dataset with the 
lowest average number of interactions per user {``\#intrns/u"} and the lowest 
average number of interactions per item ``\#u/i" (i.e., {\#intrns/u$=27.7$}, \#u/i$=13.4$).
Books is the second most sparse dataset with {``\#intrns/u"} 35.4 and ``\#u/i" 45.0.
Children and Comics are moderately sparse with {``\#intrns/u"} 57.6 and 70.0, respectively.
ML-20M and ML-1M are relatively dense with large {``\#intrns/u"} and ``\#u/i" (e.g., ``\#u/i" of ML-20M: 726.5).
Overall, we use datasets of different sparsities in our experiments to 
comprehensively compare methods in various recommendation scenarios. 
%
%

\subsection{Experimental Settings}
\label{sec:experiments:setting}
\begin{figure}[!t]
  \centering
  \includegraphics[width=0.9\linewidth]{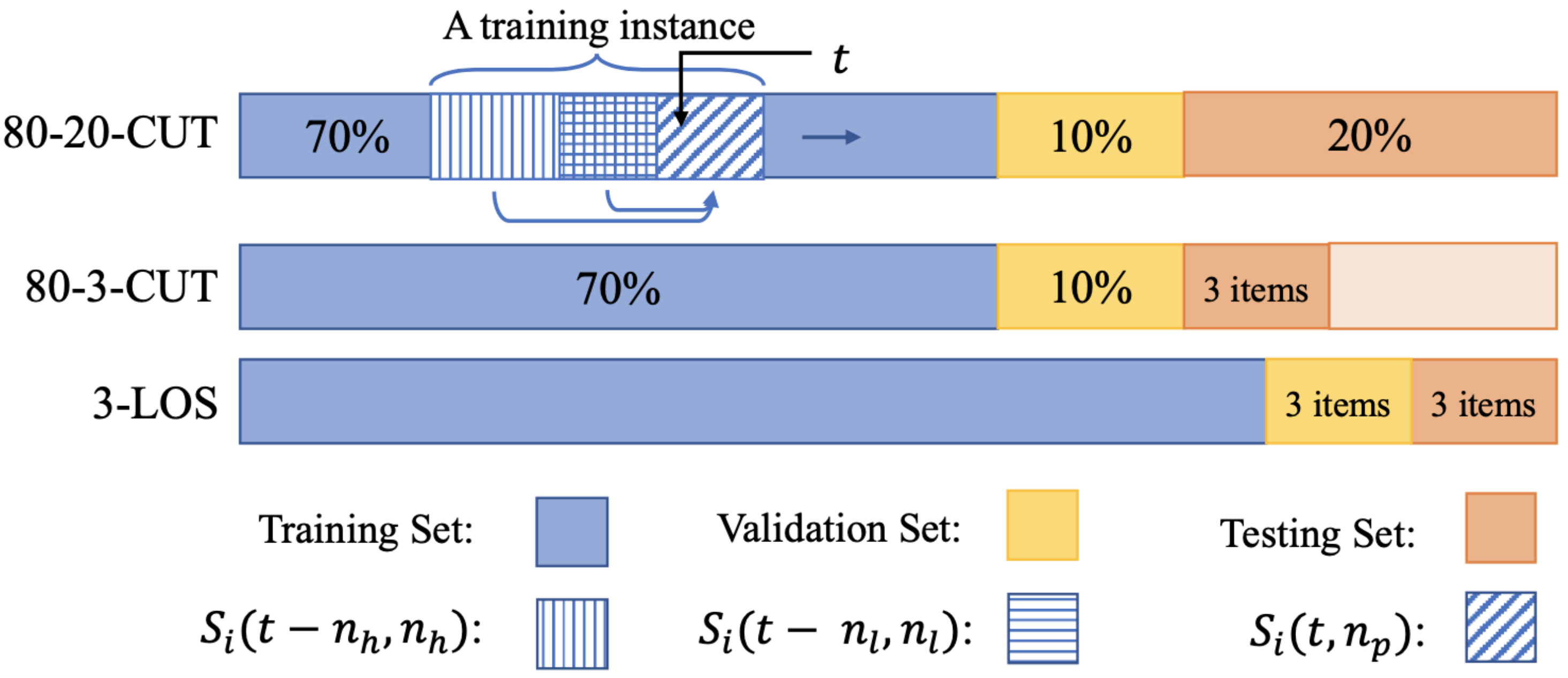}
  \caption{Experimental settings.}
  \label{fig:setting}
  \vspace{-10pt}
\end{figure}


We use three experimental settings to evaluate the methods: 
80-20-cut-off setting (\CUT, Section~\ref{sec:experiments:setting:cut}), 
80-3-cut-off setting (\CUS, Section~\ref{sec:experiments:setting:cus}) and 
leave-out setting (\LOS, Section~\ref{sec:experiments:setting:los}). 
Fig. \ref{fig:setting} presents the three settings. 
As shown in Fig.~\ref{fig:method} and Fig.~\ref{fig:setting},
in the model training, we use a sliding window of training instances 
of size $\LL$$+$$\T$
to slide item by item over the training set.
In each window, the first $\LL$ items are used to generate recommendations and
the subsequent $\T$ items are used to calculate the recommendation errors.
We update model parameters after calculating the recommendation errors 
for a batch of training instances.
In Section~\ref{sec:discussion:settings}, we will discuss 
the advantages and disadvantages of these experimental settings. 

\subsubsection{{80-20-cut-off setting (\CUT)}}
\label{sec:experiments:setting:cut}
We extract the first 70\% of each user's sequence as training set, 
the next 10\% as validation set 
for parameter tuning, and the remaining 20\% as testing  
set\footnote{We used the data splits from \url{https://github.com/allenjack/HGN}.}. 
%
During training, we train models using the training set and 
evaluate the models on the validation sets every 20 epochs. 
We select the best parameters with respect to Recall@10 values on the validation sets and use these 
parameters for testing. 
%
During testing, following \HGN and \Caser, we use both 
training and validation sets and the best parameters identified as above 
to train new models, and evaluate them on all the 
items in the testing set.
Specifically, for each user, we evaluate whether his/her items in the testing set 
are ranked as the top-$k$ items in the recommendation list using the metrics as described
in Section~\ref{sec:experiments:metrics}.
Note that if the user has a long sequence of items, his/her testing set could be large. 
\CUT is the most widely used experimental setting in sequential 
recommendation 
literature~\cite{yuan2014graph,zhao2016stellar,tang2018personalized,ma2019hierarchical}. 

\subsubsection{{80-3-cut-off setting (\CUS)}}
\label{sec:experiments:setting:cus}

We use the same training and validation set 
as in \CUT, but only the next 3 items after the validation set  
as the testing set.
The training, validation and testing processes are identical to those in \CUT, respectively. 
Compared to \CUT, \CUS recommends the immediate next 
few items, not potentially many items that might be only 
purchased/rated much later (e.g., in \CUT, 20\% of a long user sequence may have 
many and late items). 
In addition, \CUS evaluates the recommendation performance on a same number of next items (i.e., 3)
for all the users. 

\subsubsection{{Leave-3-out setting (\LOS)}}
\label{sec:experiments:setting:los}
We use only the last 3 items in each 
user sequence for testing and all the previous items for training and validation. 
The validation set contains only the 3 items before the testing items.
The training, validation and test processes are also the same as in \CUT.
%
%
\LOS maximizes the data for training and recommends the immediate 
next few items. 

\subsection{Evaluation Metrics}
\label{sec:experiments:metrics}
%
Following the literature~\cite{ma2019hierarchical}, 
we use {Recall@$k$} and {NDCG@$k$} to evaluate the different methods.  
For each user, {Recall@$k$} measures the proportion 
of all the ground-truth purchased/rated items in the testing set that are correctly recommended
among top-$k$ recommended items. 
The overall {Recall@$k$} value 
is calculated as the average over all the users. 
Higher {Recall@$k$} indicates better performance. 

{NDCG@$k$} is the normalized discounted cumulative gain for among top-$k$ ranking, 
in which $\text{gain} \in\{0,1\}$, indicating whether a ground-truth purchased/rated item has been 
recommended (i.e., 1) or not (i.e., 0). 
Thus, {NDCG@$k$} incorporates the positions of the correctly recommended 
items among the top-$k$ recommendations. 
Higher {NDCG@$k$} indicates better performance.

\section{Experimental Results}
\label{sec:results}



\subsection{Overall Performance in the \CUT Setting}
\label{sec:results:lof}




Table~\ref{tbl:performance} and Table~\ref{tbl:performanceNDCG} present
the results in terms of Recall@$5$ and Recall@$10$, NDCG@$5$ and NDCG@$10$, respectively,
on all the six datasets in the \CUT setting (Section~\ref{sec:experiments:setting:cut}).
Note that \CUT is the setting used in \Caser and \HGN.
In \CUT, we tuned all the parameters exhaustively using grid search over the validation sets.
Following the parameter tuning process as in \HGN, we evaluate the models 
on the validation sets every 20 epochs and select the best parameters with respect 
to Recall@$10$ values on the validation sets.
We report the results achieved by the selected best parameters 
on the testing sets in the tables.
In particular, in \method, we used a larger parameter space
and tuned more parameters (e.g., $d$, $\LL$, $\T$) 
than those reported~\cite{ma2019hierarchical}.
As a result, the performance in Table~\ref{tbl:performance} and Table~\ref{tbl:performanceNDCG} 
are in general better than those reported in the original \HGN paper~\cite{ma2019hierarchical} for all the baseline methods.

\begin{table}[t!]
\footnotesize
  \caption{\mbox{Overall Performance in \CUT (Recall)}} 
  \centering
  \label{tbl:performance}
  \begin{threeparttable}
      \begin{tabular}{
        @{\hspace{0pt}}l@{\hspace{1pt}}
        @{\hspace{2pt}}l@{\hspace{1pt}}
        @{\hspace{3pt}}r@{\hspace{0pt}}
        @{\hspace{1pt}}r@{\hspace{1pt}}
        @{\hspace{3pt}}r@{\hspace{1pt}}
        @{\hspace{3pt}}r@{\hspace{1pt}}
        @{\hspace{3pt}}r@{\hspace{1pt}}
        @{\hspace{3pt}}r@{\hspace{3pt}}
        @{\hspace{3pt}}r@{\hspace{1pt}}
        @{\hspace{1pt}}r@{\hspace{0pt}}
        }
        \toprule
         & Dataset & {\Caser~} & \SASRec & \HGN & \methodMax & \methodMean & \methodCorx & \methodCorm & imp\%\\
        \midrule
         \multirow{6}{*}{\rotatebox[origin=c]{90}{{\centering Recall@5}}}
         & CDs      & 0.0237 & \underline{0.0356} & 0.0337 & 0.0384 & \textbf{$\mathclap{^{\dagger~}}$0.0401} & 0.0360 & \textbf{0.0397} & 12.6$^*$\\
         & Books    & 0.0231 & \textbf{$\mathclap{^{\dagger~}}$0.0433} & 0.0317 & 0.0380 & 0.0380 & 0.0373 & \textbf{0.0412} & -4.8$^*$\\
         & Children & 0.0757 & 0.0806 & \underline{0.0866} & 0.0897 & \textbf{0.0905} & 0.0897 & \textbf{$\mathclap{^{\dagger~}}$0.0921} & 6.4$^*$\\
         & Comics   & 0.1169 & 0.1344 & \textbf{0.1355} & 0.1330 & 0.1354 & 0.1347 & \textbf{$\mathclap{^{\dagger~}}$0.1385} & 2.2$^*$\\
         & ML-20M   & 0.0781 & \underline{0.0784} & 0.0751 & 0.0751 & \textbf{0.0816} & 0.0754 & \textbf{$\mathclap{^{\dagger~}}$0.0838} & 6.9$^*$\\
         & ML-1M    & \textbf{$\mathclap{^{\dagger~}}$0.0823} & 0.0713 & 0.0727 & 0.0739 & 0.0759 & 0.0741 & \textbf{0.0793} & -3.6$^*$\\
        \midrule
 \multirow{6}{*}{\rotatebox[origin=c]{90}{{\centering Recall@10}}}
        & CDs      & 0.0381 & \underline{0.0567} & 0.0536 & 0.0598 & \textbf{$\mathclap{^{\dagger~}}$0.0621} & 0.0564 & \textbf{0.0615} & 9.5$^*$\\
        & Books    & 0.0383 & \textbf{$\mathclap{^{\dagger~}}$0.0654} & 0.0496 & 0.0574 & 0.0587 & 0.0572 & \textbf{0.0630} & -3.7$^*$\\
        & Children & 0.1174 & 0.1230 & \underline{0.1322} & 0.1361 & \textbf{0.1390} & 0.1361 & \textbf{$\mathclap{^{\dagger~}}$0.1393} & 5.4$^*$\\
        & Comics   & 0.1662 & 0.1890 & \underline{0.1900} & 0.1898 & \textbf{0.1906} & 0.1881 & \textbf{$\mathclap{^{\dagger~}}$0.1945} & 2.4$^*$\\
        & ML-20M   & \underline{0.1340} & 0.1337 & 0.1271 & 0.1259 & \textbf{0.1357} & 0.1269 & \textbf{$\mathclap{^{\dagger~}}$0.1389} & 3.7$^*$\\
        & ML-1M    & \textbf{$\mathclap{^{\dagger~}}$0.1379} & 0.1253 & 0.1252 & 0.1266 & 0.1287 & 0.1254 & \textbf{0.1330} & -3.6$^*$\\
        \bottomrule
      \end{tabular}
      \begin{tablenotes}[normal,flushleft]
      \begin{scriptsize}
      \item
          The best performance in each dataset is \textbf{~$\mathclap{^{\dagger~}}$bold}.
          The second best performance in each dataset is \textbf{bold}.
          The column ``imp\%'' presents the percentage improvement of best performance of
          \method-based methods over the 
          best performance of non-\method methods (\underline{underlined} or \textbf{bold}) in each row.
          The $^*$ indicates that the improvement is statistically significant at 95\% confidence level.
          
          \par
      \end{scriptsize}
      \end{tablenotes}
  \end{threeparttable}
  \vspace{-12pt}
\end{table}

%

%
Table~\ref{tbl:performance} and Table~\ref{tbl:performanceNDCG} together show that
overall, \methodCorm is the best performing method in \CUT.
In terms of both recall and NDCG, 
\methodCorm achieves the best performance on three datasets: Children, Comics and ML-20M,
and the second best performance on the rest three datasets: CDs, Books and ML-1M.
%
\methodMean is the second best performing method in \CUT.
In terms of both recall and NDCG, 
\methodMean achieves the best performance on the CDs dataset,
and the second best performance or (near) the second best performance on
Children, Comics and ML-20M datasets.
%
\method-based methods (\methodCorm and \methodMean) consistently outperform 
the best baseline methods on four out of the six datasets (except on Books and ML-1M, which will be discussed 
later in Section~\ref{sec:results:lof:sasrec}).
On these four datasets, the average improvement from \method-based methods over the 
best baseline methods 
is 7.0\%, 5.3\%, 6.6\%, 6.1\% 
in terms of Recall@5, Recall@10, NDCG@5, and NDCG@10, respectively.  


%

\subsubsection{Comparison between \method and \methodCor}
\label{sec:results:lof:chh}

Table~\ref{tbl:performance} and Table~\ref{tbl:performanceNDCG} together show that
\methodCorm outperforms \methodMean on all the datasets except the most sparse dataset CDs (Table~\ref{tbl:dataset})
in terms of all the evaluation metrics. 
The difference between \methodCorm and \methodMean is that \methodCorm incorporates the information 
from item synergies, whereas \methodMean doesn't explicitly model item synergies.
The superior performance of \methodCorm over \methodMean on most benchmark datasets 
indicates that 
the synergies among items may provide additional information compared to individual items, and thus could improve 
the recommendation performance.
\methodCorm achieves very similar performance with \methodMean on the CDs dataset.
This might be due to the fact that the CDs dataset is extremely sparse.
Thus, very limited information about item synergies can be learned from the data.
Table~\ref{tbl:performance} and Table~\ref{tbl:performanceNDCG} together also show that
\methodCorx always achieve very similar performance with \methodMax on all the metrics over 
the six datasets.
This indicates that when using max pooling to calculate the high-order associations, 
the latent cross mechanism could not effectively 
combine information from the high-order associations and the item synergies .
We leave the investigation of this problem in our future work.

\subsubsection{Comparison between mean pooling and max pooling}
\label{sec:results:lof:cmm}

Table~\ref{tbl:performance} and ~\ref{tbl:performanceNDCG} also 
show that \methodMean and \methodCorm consistently outperforms \methodMax and \methodCorx, respectively, 
over all the six datasets.
This indicates that mean pooling is more effective than 
max pooling on calculating the high/low-order associations
on recommendation datasets.
%
%
This may be because of that max pooling may focus on a few items with 
high values in latent features and thus fails 
to effectively capture the high-order associations 
that involve multiple items.

\begin{table}[t!]
\footnotesize
  \caption{\mbox{Overall Performance in \CUT (NDCG)}} 
  \centering
  \label{tbl:performanceNDCG}
  \begin{threeparttable}
      \begin{tabular}{
        @{\hspace{0pt}}l@{\hspace{1pt}}
        @{\hspace{2pt}}l@{\hspace{1pt}}
        @{\hspace{3pt}}r@{\hspace{0pt}}
        @{\hspace{1pt}}r@{\hspace{1pt}}
        @{\hspace{3pt}}r@{\hspace{1pt}}
        @{\hspace{3pt}}r@{\hspace{1pt}}
        @{\hspace{3pt}}r@{\hspace{1pt}}
        @{\hspace{3pt}}r@{\hspace{3pt}}
        @{\hspace{3pt}}r@{\hspace{1pt}}
        @{\hspace{1pt}}r@{\hspace{0pt}}
        }
        \toprule
         & Dataset & {\Caser~} & \SASRec & \HGN & \methodMax & \methodMean & \methodCorx & \methodCorm & imp\%\\
        \midrule
         \multirow{6}{*}{\rotatebox[origin=c]{90}{{\centering NDCG@5}}}
         & CDs      & 0.0277 & 0.0392 & \underline{0.0401} & 0.0440 & \textbf{$\mathclap{^{\dagger~}}$0.0458} & 0.0412 & \textbf{0.0452} & 14.2$^*$\\
         & Books    & 0.0345 & \textbf{$\mathclap{^{\dagger~}}$0.0614} & 0.0467 & 0.0549 & 0.0547 & 0.0551 & \textbf{0.0606} & -1.3$^*$\\
         & Children & 0.1254 & 0.1330 & \underline{0.1425} & 0.1476 & \textbf{0.1479} & 0.1475 & \textbf{$\mathclap{^{\dagger~}}$0.1523} & 6.9$^*$\\
         & Comics   & 0.2283 & \textbf{0.2631} & 0.2585 & 0.2527 & 0.2572 & 0.2587 & \textbf{$\mathclap{^{\dagger~}}$0.2662} & 1.2\textcolor{white}{$^*$}\\
         & ML-20M   & 0.1473 & \underline{0.1489} & 0.1381 & 0.1350 & \textbf{0.1496} & 0.1377 &\textbf{$\mathclap{^{\dagger~}}$0.1552} & 4.2$^*$\\
         & ML-1M    & \textbf{$\mathclap{^{\dagger~}}$0.1890} & 0.1690 & 0.1742 & 0.1666 & 0.1757 & 0.1680 & \textbf{0.1863} & -1.4\textcolor{white}{$^*$}\\
        \midrule
 \multirow{6}{*}{\rotatebox[origin=c]{90}{{\centering NDCG@10}}}
        & CDs      & 0.0327 & \underline{0.0470} & 0.0469 & 0.0517 & \textbf{$\mathclap{^{\dagger~}}$0.0534} & 0.0485 & \textbf{0.0528} & 13.6$^*$\\
        & Books    & 0.0383 & \textbf{$\mathclap{^{\dagger~}}$0.0663} & 0.0505 & 0.0590 & 0.0593 & 0.0591 & \textbf{0.0649} & -2.1$^*$\\
        & Children & 0.1305 & 0.1371 & \underline{0.1467} & 0.1517 & \textbf{0.1532} & 0.1514 & \textbf{$\mathclap{^{\dagger~}}$0.1559} & 6.3$^*$\\
        & Comics   & 0.2157 & \textbf{0.2468} & 0.2436 & 0.2408 & 0.2434 & 0.2425 & \textbf{$\mathclap{^{\dagger~}}$0.2507} & 1.6\textcolor{white}{$^*$}\\
        & ML-20M   & 0.1573 & \underline{0.1587} & 0.1473 & 0.1433 & \textbf{0.1584} & 0.1467 & \textbf{$\mathclap{^{\dagger~}}$0.1635} & 3.0$^*$\\
        & ML-1M    & \textbf{$\mathclap{^{\dagger~}}$0.1904} & 0.1727 & 0.1763 & 0.1719 & 0.1778 & 0.1714 & \textbf{0.1875} & -1.5$^*$\\
        \bottomrule
      \end{tabular}
      \begin{tablenotes}[normal,flushleft]
      \begin{scriptsize}
      \item
          The columns in this table have the same meanings as those in Table~\ref{tbl:performance}.
          \par
      \end{scriptsize}
      \end{tablenotes}
  \end{threeparttable}
  \vspace{-10pt}
\end{table}


\subsubsection{Comparison \method Methods with \HGN}
\label{sec:results:lof:hgn}
%
%
%

Table~\ref{tbl:performance} and Table~\ref{tbl:performanceNDCG} show
that \method methods (\methodMean and \methodCorm) consistently outperforms \HGN on all the six datasets.
Note that \HGN is the state-of-the-art method in literature~\cite{ma2019hierarchical} and outperforms 
many other baseline methods. 
The difference between \method and \HGN is that \method uses mean pooling over the 
last $\LL$ items and last $\M$ items to recommend the next items, 
whereas \HGN uses gating mechanisms  
over the last items and also over their latent features
to differentiate the importance of these items and their features. 
However, as Table~\ref{tbl:dataset} shows, each user typically has only a few items (compared to 
all the possible items), 
and each item is typically only purchased/rated by a few users (compared to all the possible users). 
Therefore, the data sparsity issue
may lead to less meaningful gating weights learned by parameterized gating mechanisms (more discussion in 
Section~\ref{sec:discussion:weight}), 
whereas equal weights from mean pooling would suffice. 
In addition, \method combines both high-order and low-order sequential patterns, conforming 
to the discovery in~\cite{tang2018personalized}, which may also contribute to the superior performance. 
Moreover, \methodCorm explicitly models the synergies among items, which may 
provide additional information for the recommendation and  contributes to the superior performance
as well.
%

\subsubsection{Comparison \method Methods with \SASRec and \Caser}
\label{sec:results:lof:sasrec}

%
Table~\ref{tbl:performance} and Table~\ref{tbl:performanceNDCG} show that
both \methodMean and \methodCorm outperform \SASRec on 5 out of 6 datasets 
and achieve similar performance as \SASRec on the rest Books dataset.
%
A key difference between \method and \SASRec is that \method leverages item associations 
and item synergies in the 
most recent purchases/ratings, whereas \SASRec only uses the long-term user preferences. 
As demonstrated in literature~\cite{zhou2019deep}, 
user preferences may shift over time and thus preferences 
from the most recent purchases/ratings might provide more pertinent information 
for the next recommendations.
In addition, similar to \HGN, \SASRec adapts attention mechanisms 
to identify most informative items. 
However, as discussed in Section~\ref{sec:results:lof:hgn}, 
the data sparsity issue may result in less meaningful weights, whereas equal weights from mean pooling 
may lead to better performance.
We notice that \methodCorm achieves slightly worse performance compared to that of \SASRec 
on the Books dataset (\methodCorm: 0.0412 vs \SASRec: 0.0433 on Recall@$5$, 
\methodCorm: 0.0630 vs \SASRec: 0.0654 on Recall@$10$).
This might be due to the reason that users have in general strong long-term 
preferences on books.
Thus, \SASRec, which focuses on modeling the long-term user preferences, 
outperforms \methodCorm, which learns users' preferences primarily from users' 
most recent purchases/ratings on the Books dataset.
We also notice that on Comics, the improvement of \methodCorm over \SASRec is 
statistically significant 
on Recall@$5$ and Recall@$10$, while not statistically significant on NDCG@$5$ and NDCG@$10$.
This indicates that on Comics, \methodCorm works significantly better 
than \SASRec on recommending the items of users' interest (i.e., Recall), 
and slightly better than \SASRec on ranking the items of users' interest 
on top (i.e., NDCG).
%
%
%
In addition, Table~{\ref{tbl:performance}} and Table~{\ref{tbl:performanceNDCG}} show 
that \methodMean and \methodCorm outperform \Caser except on the  
relatively dense dataset ML-1M, 
%
which may exhibit strong local patterns in the latent item features that \Caser 
could well capture utilize.
However, in terms of NDCG@$5$, the improvement of \Caser over \methodCorm on ML-1M is not statistically significant, 
indicating that \methodCorm could still achieve very similar performance with \Caser on ML-1M.

%

\subsubsection{Summary}
\label{sec:results:lof:sum}

%
Overall, \method outperforms the best baseline methods with very high percentage improvement 
when the dataset is very sparse (e.g., 9.5\% on CDs on Recall@$10$), 
and significant improvement 
when the datasets is moderately sparse (e.g., 5.4\% on Children on Recall@$10$). 
When the datasets are dense (e.g., ML-1M) 
or 
in the datasets, users' behaviors are significantly affected by users' long-term preferences, 
\method could be slightly worse 
than the other baseline methods. However, most of the sequential recommendation problems 
always have very sparse and time sensitive
datasets, on which \method will be effective. 
More detailed analysis at the data sparsity aspect 
is available in Section~\ref{sec:discussion:weight}.

\subsection{Overall Performance in the \CUS Setting}
\label{sec:results:lot}

%
\begin{table}[!t]
\footnotesize
  \caption{\mbox{Overall Performance in \CUS (Recall)}}
  \centering
  \label{tbl:near}
  \begin{threeparttable}
      \begin{tabular}{
        @{\hspace{0pt}}l@{\hspace{1pt}}
        @{\hspace{2pt}}l@{\hspace{1pt}}
        @{\hspace{3pt}}r@{\hspace{0pt}}
        @{\hspace{1pt}}r@{\hspace{1pt}}
        @{\hspace{3pt}}r@{\hspace{1pt}}
        @{\hspace{3pt}}r@{\hspace{1pt}}
        @{\hspace{3pt}}r@{\hspace{1pt}}
        @{\hspace{3pt}}r@{\hspace{3pt}}
        @{\hspace{3pt}}r@{\hspace{1pt}}
        @{\hspace{1pt}}r@{\hspace{0pt}}
        }
        \toprule
        & Dataset & {\Caser~} & \SASRec & \HGN & \methodMax & \methodMean & \methodCorx & \methodCorm & imp\%\\
        \midrule
        \multirow{6}{*}{\rotatebox[origin=c]{90}{{\centering Recall@5}}}
        & CDs      & 0.0271 & \underline{0.0398} & 0.0395 & 0.0433 & \textbf{$\mathclap{^{\dagger~}}$0.0449} & 0.0423 & \textbf{0.0442} & 12.8$^*$\\
        & Books    & 0.0306 & \textbf{$\mathclap{^{\dagger~}}$0.0589} & 0.0438 & 0.0517 & 0.0511 &  0.0512 & \textbf{0.0555} & -5.8$^*$\\
        & Children & 0.1001 & 0.1360 & \underline{0.1404} & \textbf{0.1446} & 0.1436 & 0.1432 & \textbf{$\mathclap{^{\dagger~}}$0.1483} & 5.6$^*$\\
        & Comics   & 0.1510 & \textbf{$\mathclap{^{\dagger~}}$0.2689} & 0.2484 & 0.2454 & 0.2528 & 0.2529 & \textbf{0.2563} & -4.7$^*$\\
        & ML-20M   & \underline{0.1119} & 0.1097 & 0.1022 & 0.1032 & \textbf{0.1133} & 0.1031 & \textbf{$\mathclap{^{\dagger~}}$0.1166} & 4.2$^*$\\
        & ML-1M    & 0.1209 & \underline{0.1293} & 0.1243 & 0.1274 & \textbf{0.1307} & 0.1248 & \textbf{$\mathclap{^{\dagger~}}$0.1364} & 5.5$^*$\\
        \midrule
        \multirow{6}{*}{\rotatebox[origin=c]{90}{{\centering Recall@10}}}
        & CDs      & 0.0438 & \underline{0.0624} & 0.0616 & 0.0659 & \textbf{$\mathclap{^{\dagger~}}$0.0696} & 0.0638 & \textbf{0.0687} & 11.5$^*$\\
        & Books    & 0.0400 & \textbf{$\mathclap{^{\dagger~}}$0.0869} & 0.0671 & 0.0774 & 0.0777 & 0.0766 & \textbf{0.0835} & -3.9$^*$\\
        & Children & 0.1589 & 0.1939 & \underline{0.2013} & 0.2063 & \textbf{0.2078} & 0.2054 & \textbf{$\mathclap{^{\dagger~}}$0.2111} & 4.9$^*$\\
        & Comics   & 0.2102 & \textbf{$\mathclap{^{\dagger~}}$0.3372} & 0.3188 & 0.3190 & 0.3243 & 0.3220 & \textbf{0.3278} & -2.8$^*$\\
        & ML-20M   & \textbf{0.1858} & 0.1839 & 0.1685 & 0.1678 & 0.1825 & 0.1687 & \textbf{$\mathclap{^{\dagger~}}$0.1878} & 1.1$^*$\\
        & ML-1M    & 0.1933 & \underline{0.2113} & 0.2065 & 0.2077 & \textbf{0.2145} & 0.2074 & \textbf{$\mathclap{^{\dagger~}}$0.2221} & 5.1$^*$\\
        \bottomrule
      \end{tabular}
      \begin{tablenotes}[normal,flushleft]
      \begin{scriptsize}
      \item
          The columns in this table have the same meanings as those in Table~\ref{tbl:performance}.
          \par
      \end{scriptsize}
      \end{tablenotes}
  \end{threeparttable}
  \vspace{-10pt}
\end{table}


Table~\ref{tbl:near} and Table~\ref{tbl:nearNDCG} present the results of different methods 
in the \CUS setting (Section~\ref{sec:experiments:setting:cus}).
In \CUS, for \method methods and all the baseline methods, 
the parameters are also tuned by grid search on the validation sets.
The parameters tuning protocol is the same as in \CUT.
%
Overall, the performance in \CUS of \method methods and the baseline methods 
has a similar trend as that in \CUT. 
%
Particularly, \methodCorm is still the best performing method in this setting.
In terms of Recall@$5$ and Recall@$10$, 
\methodCorm achieves the best performance on 3 out of 6 datasets 
(i.e., Children, ML-20M and ML-1M), 
and the second best performance on the rest CDs, Books and Comics datasets. 
In terms of NDCG@$5$ and NDCG@$10$, \methodCorm 
also achieves the best performance on the Children, ML-20M and ML-1M datasets.
On the rest CDs, Books and Comics datasets, \methodCorm achieves very similar performance 
compared to the best method on these datasets.
For example, on the CDs dataset, \methodMean 
achieves the best Recall@$5$ 0.0449 and 
\methodCorm achieves the second best Recall@$5$ 0.0442 (i.e., 1.6\% difference)
%

Table~\ref{tbl:near} and Table~\ref{tbl:nearNDCG} also show that 
\methodCorm doesn't work as well as \SASRec on the Books and Comics datasets 
(e.g., \methodCorm: 0.0555 and \SASRec: 0.0589 on Recall@$5$ on Books).
On Books, as discussed in Section~\ref{sec:results:lof:sasrec}, 
there could exist strong long-term preferences among users and thus 
\SASRec could work well.
As shown in Table~\ref{tbl:dataset}, 
Comics is much denser than CDs and Books.
Thus, the attention mechanism in \SASRec could learn meaningful weights and 
improve the recommendation performance.
%
%
Even though \methodCorm does not outperform \SASRec on Books and Comics,
it still achieves the second best performance on these two datasets, and 
statistically significantly outperforms \SASRec on the other datasets (e.g., CDs).

\begin{table}[t!]
\footnotesize
  \caption{\mbox{Overall Performance in \CUS (NDCG)}}
  \centering
  \label{tbl:nearNDCG}
  \begin{threeparttable}
      \begin{tabular}{
        @{\hspace{0pt}}l@{\hspace{1pt}}
        @{\hspace{2pt}}l@{\hspace{1pt}}
        @{\hspace{3pt}}r@{\hspace{0pt}}
        @{\hspace{1pt}}r@{\hspace{1pt}}
        @{\hspace{3pt}}r@{\hspace{1pt}}
        @{\hspace{3pt}}r@{\hspace{1pt}}
        @{\hspace{3pt}}r@{\hspace{1pt}}
        @{\hspace{3pt}}r@{\hspace{3pt}}
        @{\hspace{3pt}}r@{\hspace{1pt}}
        @{\hspace{1pt}}r@{\hspace{0pt}}
        }
        \toprule
        & Dataset & {\Caser~} & \SASRec & \HGN & \methodMax & \methodMean & \methodCorx & \methodCorm & imp\%\\
        \midrule
        \multirow{6}{*}{\rotatebox[origin=c]{90}{{\centering NDCG@5}}}
        & CDs      & 0.0239 & \underline{0.0357} & 0.0354 & 0.0389 & \textbf{$\mathclap{^{\dagger~}}$0.0403} & 0.0378 & \textbf{0.0400} & 12.9$^*$\\
        & Books    & 0.0274 & \textbf{$\mathclap{^{\dagger~}}$0.0554} & 0.0403 & 0.0489 & 0.0476 & 0.0474 & \textbf{0.0518} & -6.5$^*$\\
        & Children & 0.0890 & 0.1263 & \underline{0.1281} & \textbf{0.1323} & 0.1314 & 0.1317 & \textbf{$\mathclap{^{\dagger~}}$0.1355} & 5.8$^*$\\
        & Comics   & 0.1376 & \textbf{$\mathclap{^{\dagger~}}$0.2615} & 0.2302 & 0.2273 & 0.2356 & 0.2351 & \textbf{0.2398} & -8.3$^*$\\
        & ML-20M   & \underline{0.0989} & 0.0963 & 0.0904 & 0.0921 & \textbf{0.1006} & 0.0915 & \textbf{$\mathclap{^{\dagger~}}$0.1044} & 5.6$^*$\\
        & ML-1M    & 0.1077 & \underline{0.1128} & 0.1083 & 0.1109 & \textbf{0.1172} & 0.1110 & \textbf{$\mathclap{^{\dagger~}}$0.1221} & 8.2$^*$\\
        \midrule
        \multirow{6}{*}{\rotatebox[origin=c]{90}{{\centering NDCG@10}}}
        & CDs      & 0.0313 & \underline{0.0458} & 0.0452 & 0.0489 & \textbf{$\mathclap{^{\dagger~}}$0.0514} & 0.0474 & \textbf{0.0509} & 12.2$^*$\\
        & Books    & 0.0361 & \textbf{$\mathclap{^{\dagger~}}$0.0680} & 0.0508 & 0.0604 & 0.0597 & 0.0590 & \textbf{0.0644} & -5.3$^*$\\
        & Children & 0.1156 & 0.1526 & \underline{0.1556} & 0.1603 & \textbf{0.1605} & 0.1599 & \textbf{$\mathclap{^{\dagger~}}$0.1640} & 5.4$^*$\\
        & Comics   & 0.1645 & \textbf{$\mathclap{^{\dagger~}}$0.2927} & 0.2623 & 0.2609 & 0.2683 & 0.2666 & \textbf{0.2724} & -6.9$^*$\\
        & ML-20M   & \textbf{0.1322} & 0.1297 & 0.1203 & 0.1212 & 0.1321 & 0.1211 & \textbf{$\mathclap{^{\dagger~}}$0.1365} & 3.3$^*$\\
        & ML-1M    & 0.1405 & \underline{0.1498} & 0.1454 & 0.1486 & \textbf{0.1550} & 0.1483 & \textbf{$\mathclap{^{\dagger~}}$0.1608} & 7.3$^*$\\
        \bottomrule
      \end{tabular}
      \begin{tablenotes}[normal,flushleft]
      \begin{scriptsize}
      \item
          The columns in this table have the same meanings as those in Table~\ref{tbl:performance}.
          \par
      \end{scriptsize}
      \end{tablenotes}
  \end{threeparttable}
  \vspace{-10pt}
\end{table}


\subsubsection{Comparison with the \CUT Setting}
\label{sec:results:lot:complof}
%
%
%
%
A comparison between Table~\ref{tbl:performance} (in the \CUT setting) and Table~\ref{tbl:near} (in the \CUS setting) 
shows that in terms of Recall@$5$ and Recall@$10$, 
all the methods have in general better performance in \CUS 
(evaluating on the immediate next few items) 
than in \CUT (evaluating on the rest 20\% items). 
Note that \CUS and \CUT have the same training sets but different testing sets (Fig.~\ref{fig:setting})
%
%
This might be due to that as there are many testing items in \CUT, and therefore, 
when calculating recall values on the testing set, 
we will have a large denominator (i.e., the number of testing items)
leading to low recall values. 

%

However, the trend for NDCG@$5$ and NDCG@$10$ is different from 
that for Recall@$5$ and Recall@$10$.
A comparison between Table~\ref{tbl:performanceNDCG} and~\ref{tbl:nearNDCG} shows that 
in terms of NDCG@$5$ and NDCG@$10$, 
all the methods have in general worse performance in {\CUS}
than in {\CUT}.
%
This might be due to that when calculating NDCG values, 
the denominator will not change with the number of testing items.
However, as there are many testing items in \CUT, the recommended 
items are more likely to be found in the testing set, and thus the numerator 
could be large and result in high NDCG values in \CUT.
%
We will discuss this issue in detail later in Section~\ref{sec:discussion:settings}.
%
In addition, it is worthy noting that although \methodCorm 
underperforms \Caser on the ML-1M dataset in \CUT, 
it outperforms \Caser in \CUS.
It indicates that \method can outperform \Caser 
in recommending the immediate next few items on the relatively dense datasets.


\subsection{Overall Performance in the \LOS Setting}
\label{sec:results:loo}

\begin{table}[t!]
\footnotesize
  \caption{\mbox{Overall Performance in \LOS (Recall)}}
  \centering
  \label{tbl:performancelevel}
  \begin{threeparttable}
      \begin{tabular}{
        @{\hspace{0pt}}l@{\hspace{1pt}}
        @{\hspace{2pt}}l@{\hspace{1pt}}
        @{\hspace{3pt}}r@{\hspace{0pt}}
        @{\hspace{1pt}}r@{\hspace{1pt}}
        @{\hspace{3pt}}r@{\hspace{1pt}}
        @{\hspace{3pt}}r@{\hspace{1pt}}
        @{\hspace{3pt}}r@{\hspace{1pt}}
        @{\hspace{3pt}}r@{\hspace{3pt}}
        @{\hspace{3pt}}r@{\hspace{1pt}}
        @{\hspace{1pt}}r@{\hspace{0pt}}
        }
        \toprule
        & Dataset & {\Caser~} & \SASRec & \HGN & \methodMax & \methodMean & \methodCorx & \methodCorm & imp\%\\
        \midrule
        \multirow{6}{*}{\rotatebox[origin=c]{90}{{\centering Recall@5}}}
        & CDs      & 0.0289 & 0.0385 & \underline{0.0389} & 0.0401 & \textbf{$\mathclap{^{\dagger~}}$0.0437} & 0.0364 & \textbf{0.0430} & 12.3${^*}$\\
        & Books    & 0.0264 & \textbf{$\mathclap{^{\dagger~}}$0.0653} & 0.0440 & 0.0483 & 0.0520 & 0.0495 & \textbf{0.0528} & -19.1$^*$\\
        & Children & 0.0860 & \underline{0.1055} & 0.1021 & 0.1069 & \textbf{0.1100} & 0.1039 & \textbf{$\mathclap{^{\dagger~}}$0.1110} & 5.2$^*$\\
        & Comics   & 0.1726 & 0.0374 & \underline{0.1933} & 0.1933 & \textbf{0.1957} & 0.1960 & \textbf{$\mathclap{^{\dagger~}}$0.2033} &5.2$^*$\\
        & ML-20M   & 0.0963 & \textbf{0.0998} & 0.0899 & 0.0910 & 0.0982 & 0.0889 & \textbf{$\mathclap{^{\dagger~}}$0.1002} & 0.4\textcolor{white}{$^*$}\\
        & ML-1M    & \textbf{0.1191} & 0.1095 & 0.1119 & 0.1097 & 0.1148 & 0.1095 & \textbf{$\mathclap{^{\dagger~}}$0.1198} & 0.6\textcolor{white}{$^*$}\\
        \midrule
        \multirow{6}{*}{\rotatebox[origin=c]{90}{{\centering Recall@10}}}
        & CDs      & 0.0456 & \underline{0.0621} & 0.0600 & 0.0629 & \textbf{$\mathclap{^{\dagger~}}$0.0669} & 0.0577 & \textbf{0.0662} & 7.7$^*$\\
        & Books    & 0.0437 & \textbf{$\mathclap{^{\dagger~}}$0.0958} & 0.0665 & 0.0725 & 0.0783 & 0.0750 & \textbf{0.0798} & -16.7$^*$\\
        & Children & 0.1280 & \underline{0.1503} & 0.1497 & 0.1544 & \textbf{0.1575} & 0.1503 & \textbf{$\mathclap{^{\dagger~}}$0.1588} & 5.7$^*$\\
        & Comics   & 0.2198 & 0.0597 & \underline{0.2445} & 0.2460 & \textbf{0.2484} & 0.2470 & \textbf{$\mathclap{^{\dagger~}}$0.2564} & 4.9$^*$\\
        & ML-20M   & 0.1591 & \textbf{$\mathclap{^{\dagger~}}$0.1660} & 0.1487 & 0.1473 & 0.1586 & 0.1455 & \textbf{0.1612} & -2.9$^*$\\
        & ML-1M    & \textbf{$\mathclap{^{\dagger~}}$0.1910} & 0.1822 & 0.1790 & 0.1794 & 0.1854 & 0.1779 & \textbf{0.1898} & -0.6\textcolor{white}{$^*$}\\
        \bottomrule
      \end{tabular}
      \begin{tablenotes}[normal,flushleft]
      \begin{scriptsize}
      \item
          The columns in this table have the same meanings as those in Table~\ref{tbl:performance}.
          \par
      \end{scriptsize}
      \end{tablenotes}
  \end{threeparttable}
  \vspace{-10pt}
\end{table}


Table~\ref{tbl:performancelevel} and Table~\ref{tbl:performancelevelNDCG} present the 
results of the different methods in the \LOS setting (Section~\ref{sec:experiments:setting:los}).
Overall, \method methods (i.e., \methodMax, \methodMean, \methodCorx and \methodCorm) 
still outperform the other methods.
Particularly, \methodCorm is still the best performing method in \LOS.
In terms of Recall@$5$, \methodCorm achieves the best performance on 4 out of 6 datasets 
and the second best performance on the rest CDs and Books datasets.
In terms of NDCG@$5$, \methodCorm achieves the best performance on 5 out of 6 datasets and 
the second best performance on Books.
%
%

Table~\ref{tbl:performancelevel} and~\ref{tbl:performancelevelNDCG} also show that 
\methodCorm doesn't perform as well as \SASRec on the Books and ML-20M datasets. 
Especially on the Books dataset, \methodCorm underperforms \SASRec by 19.1\% and 20.5\% 
on Recall@$5$ and NDCG@$5$, respectively.
The good performance of \SASRec in \LOS might 
be due to that in \LOS (training on all the items except the 6 items used 
for validation and testing),
the data sparsity issue is less severe than that in \CUT and \CUS (training on 70\% items).
Thus the attention mechanism used in \SASRec may learn meaningful weights and 
thus, on some datasets, \SASRec is able to achieve better performance than that of \methodCorm.  
Even though \methodCorm works worse than \SASRec on Books and ML-20M, it still achieves 
the second best performance.
%
%

Please recall that we did a very comprehensive parameter study for 
\SASRec in order to identify its best performance. 
During the parameter study, 
we observed that \SASRec could be very sensitive to the parameters.
For example, in our experiments on the Comics dataset, 
\SASRec can achieve good performance (e.g., 0.3353 on Recall@$10$) 
on its validation set via a set of parameters.
However, by slightly changing the parameters,
the performance can decrease dramatically (e.g., from 0.3353 Recall@$10$ to 0.0587, 
that is, 6-fold decrease) on the validation set.
This issue may prevent \SASRec to perform well on the testing set because
the optimal parameters on the validation set may lead to dramatically 
different performance on the testing set. 
%
%
%
Compared to \SASRec, \method is much more stable and less sensitive to parameters as will 
be discussed in Section~\ref{sec:results:parameter}. 
More details of the parameter sensitivity issue of \SASRec
are described in Section~\ref{sec:sasrec:parameter:comics} in the Appendix.
We also observed that \SASRec requires a large amount of memory during training and 
some parameters could cause the out-of-memory issue.
The details of this issue are also described in Section~\ref{sec:sasrec:parameter:comics} 
in the Appendix.
Table~\ref{tbl:performancelevel} and~\ref{tbl:performancelevelNDCG} also 
show that \methodCorm work slightly worse than \Caser on ML-1M in terms of Recall@$10$ and NDCG@$10$.
However, the difference is not statistically significant. 
%
%
%

\begin{table}[t!]
\footnotesize
  \caption{\mbox{Overall Performance in \LOS (NDCG)}}
  \centering
  \label{tbl:performancelevelNDCG}
  \begin{threeparttable}
      \begin{tabular}{
        @{\hspace{0pt}}l@{\hspace{1pt}}
        @{\hspace{2pt}}l@{\hspace{1pt}}
        @{\hspace{3pt}}r@{\hspace{0pt}}
        @{\hspace{1pt}}r@{\hspace{1pt}}
        @{\hspace{3pt}}r@{\hspace{1pt}}
        @{\hspace{3pt}}r@{\hspace{1pt}}
        @{\hspace{3pt}}r@{\hspace{1pt}}
        @{\hspace{3pt}}r@{\hspace{3pt}}
        @{\hspace{3pt}}r@{\hspace{1pt}}
        @{\hspace{1pt}}r@{\hspace{0pt}}
        }
        \toprule
        & Dataset & {\Caser~} & \SASRec & \HGN & \methodMax & \methodMean & \methodCorx & \methodCorm & imp\%\\
        \midrule
        \multirow{6}{*}{\rotatebox[origin=c]{90}{{\centering NDCG@5}}}
        & CDs      & 0.0267 & \underline{0.0346} & 0.0342 & \textbf{0.0367} & \textbf{$\mathclap{^{\dagger~}}$0.0394} & 0.0328 & \textbf{$\mathclap{^{\dagger~}}$0.0394} & 13.9$^*$\\
        & Books    & 0.0240 & \textbf{$\mathclap{^{\dagger~}}$0.0620} & 0.0416 & 0.0456 & 0.0486 & 0.0462 & \textbf{0.0493} & -20.5$^*$\\
        & Children & 0.0790 & \textbf{0.0999} & 0.0940 & 0.0994 & \textbf{$\mathclap{^{\dagger~}}$0.1030} & 0.0966 & \textbf{$\mathclap{^{\dagger~}}$0.1030} & 3.1$^*$\\
        & Comics   & 0.1666 & 0.0344 & \underline{0.1862} & 0.1854 & \textbf{0.1881} & 0.1909 & \textbf{$\mathclap{^{\dagger~}}$0.1975} & 6.1$^*$ \\
        & ML-20M   & 0.0850 & \textbf{0.0886} & 0.0797 & 0.0815 & 0.0882 & 0.0795 & \textbf{$\mathclap{^{\dagger~}}$0.0900} & 1.6$^*$ \\
        & ML-1M    & \textbf{0.1067} & 0.0974 & 0.0996 & 0.0987 & 0.1026 & 0.0977 & \textbf{$\mathclap{^{\dagger~}}$0.1070} & 0.3\textcolor{white}{$^*$} \\
        \midrule
        \multirow{6}{*}{\rotatebox[origin=c]{90}{{\centering NDCG10}}}
        & CDs      & 0.0342 & \underline{0.0453} & 0.0444 & \textbf{0.0471} & \textbf{$\mathclap{^{\dagger~}}$0.0499} & 0.0425 & \textbf{$\mathclap{^{\dagger~}}$0.0499} & 10.2$^*$ \\
        & Books    & 0.0318 & \textbf{$\mathclap{^{\dagger~}}$0.0758} & 0.0518 & 0.0566 & 0.0605 & 0.0577 & \textbf{0.0615} & -18.9$^*$\\ 
        & Children & 0.0980 & \underline{0.1201} & 0.1155 & \textbf{0.1208} & \textbf{$\mathclap{^{\dagger~}}$0.1246} & 0.1177 & \textbf{$\mathclap{^{\dagger~}}$0.1246} & 3.7$^*$\\ 
        & Comics   & 0.1881 & 0.0445  & \underline{0.2094} & 0.2094 & \textbf{0.2121} & 0.2140 & \textbf{$\mathclap{^{\dagger~}}$0.2217} & 5.9$^*$\\
        & ML-20M   & 0.1136 & \textbf{$\mathclap{^{\dagger~}}$0.1185} & 0.1063 & 0.1069 & 0.1156 & 0.1050 & \textbf{0.1176} & -0.8$^*$\\
        & ML-1M    & \textbf{$\mathclap{^{\dagger~}}$0.1391} & 0.1302 & 0.1299 & 0.1302 & 0.1346 & 0.1286 & \textbf{0.1387} & -0.3\textcolor{white}{$^*$}\\
        \bottomrule
      \end{tabular}
      \begin{tablenotes}[normal,flushleft]
      \begin{scriptsize}
      \item
          The columns in this table have the same meanings as those in Table~\ref{tbl:performance}.
          \par
      \end{scriptsize}
      \end{tablenotes}
  \end{threeparttable}
  \vspace{-10pt}
\end{table}


%
\subsubsection{Comparison with the \CUS Setting}
\label{sec:results:los:compcus}

A comparison between Table~\ref{tbl:near} and Table~\ref{tbl:performancelevel}, and between 
Table~\ref{tbl:nearNDCG} and Table~\ref{tbl:performancelevelNDCG} 
shows that in terms 
of both Recall (i.e., Recall@$5$ and Recall@$10$) and NDCG (i.e., NDCG@$5$ and NDCG@$10$), 
all the methods have in general better performance in \CUS
(i.e., the next 3 items after the validation set of each user are used for testing and 
the first 80\% sequence are used for training and validation) 
than in \LOS (i.e., the last 3 items of each user are used for testing and 
all the previous items are used for training and validation).
Compared to \CUS setting, the training sets in \LOS setting contain more early purchases/ratings 
(i.e., purchases/ratings that occurred long time ago before the testing items).
These early purchases/ratings may not accurately represent users' preferences at the time
of the testing items as such preferences may shift~\cite{zhou2019deep}.

\subsection{Performance Summary among All the Settings}
\label{sec:results:all}
%


Table~\ref{tbl:impro} presents the average improvement of \methodCorm compared to 
\Caser, \SASRec, \HGN and \methodMean in each of the experimental settings \CUT, \CUS and \LOS.
The average improvement is calculated as the mean of the improvement of \methodCorm 
over the compared methods (i.e., \Caser, \SASRec, \HGN and \methodMean) 
over all the datasets.
%
We also present the average improvement of \methodCorm over \SASRec 
when the Books dataset is excluded (i.e., the ``(-B)" column in Table~\ref{tbl:impro}) 
due to that the strong performance of \SASRec on Books in \LOS 
(Table~\ref{tbl:performancelevel} and Table~\ref{tbl:performancelevelNDCG}) 
may dominate the average improvement. 
%
%
Considering the parameter sensitivity issue of \SASRec (Section~\ref{sec:results:loo}) in \LOS,
we exclude the Comics dataset when calculating the average improvement 
of {\methodCorm} over {\SASRec} in {\LOS} (i.e., the ``(-C)" column in Table~\ref{tbl:impro}). 
Please note that \methodMean consistently outperforms \methodMax and \methodCorx 
in all the three settings
as shown in Sections~\ref{sec:results:lof}, \ref{sec:results:lot} and~\ref{sec:results:loo}.
Thus, we compare \methodCorm only with \methodMean, not with \methodMax or \methodCorx 
that \methodMean outperforms.

\begin{table}[!t]
\footnotesize
  \caption{\mbox{{Performance Improvement of {\methodCorm} (\%)}}}
  \centering
  \label{tbl:impro}
  \begin{threeparttable}
      \begin{tabular}{
        @{\hspace{0pt}}l@{\hspace{6pt}}
        @{\hspace{6pt}}l@{\hspace{6pt}}
        @{\hspace{6pt}}r@{\hspace{6pt}}
        @{\hspace{6pt}}r@{\hspace{6pt}}
        @{\hspace{1pt}}r@{\hspace{8pt}}
        @{\hspace{6pt}}r@{\hspace{6pt}}
    @{\hspace{6pt}}r@{\hspace{0pt}}
        }
        \toprule
        \multirow{2}{*}{Setting} & \multirow{2}{*}{Metric} & \multirow{2}{*}{\Caser} 
	& \multicolumn{2}{c}{\SASRec} & \multirow{2}{*}{\HGN} & \multirow{2}{*}{\methodMean} \\
        \cmidrule(lr){4-5}
        &    	     & 	   & \scriptsize{all} & \scriptsize{-B} & &\\
        \midrule
        \multirow{4}{*}{{{\centering \CUT}}}
        & Recall@$5$  & 31.6$^*$ & 7.0$^*$ & 9.4$^*$ & 12.8$^*$ & 3.1$^*$\\
        & Recall@$10$ & 26.9$^*$ & 5.2$^*$ & 6.9$^*$ & 10.8$^*$ & 2.4$^*$\\
        & NDCG@$5$    & 30.1$^*$ & 7.4$^*$ & 9.1$^*$ & 11.9$^*$ & 4.3$^*$\\
        & NDCG@$10$   & 28.2$^*$ & 6.2$^*$ & 7.8$^*$ & 11.3$^*$ & 3.6$^*$\\
        \midrule
        \multirow{4}{*}{{{\centering \CUS}}}
        & Recall@$5$  & 46.6$^*$ & 3.6\textcolor{white}{$^*$} & 5.4\textcolor{white}{$^*$} & 11.9$^*$ & 3.2$^*$\\
        & Recall@$10$ & 45.1$^*$ & 3.2\textcolor{white}{$^*$} & 4.7\textcolor{white}{$^*$} & 10.4$^*$ & 2.5$^*$\\
        & NDCG@$5$    & 50.3$^*$ & 3.5\textcolor{white}{$^*$} & 5.5\textcolor{white}{$^*$} & 13.3$^*$ & 3.5$^*$\\
        & NDCG@$10$   & 44.4$^*$ & 3.2\textcolor{white}{$^*$} & 4.9\textcolor{white}{$^*$} & 12.1$^*$ & 2.9$^*$\\
        \midrule
        \multirow{2}{*}{Setting} & \multirow{2}{*}{Metric} & \multirow{2}{*}{\Caser} 
        & \multicolumn{2}{c}{\SASRec} & \multirow{2}{*}{\HGN} 
        & \multirow{2}{*}{\methodMean} \\
        \cmidrule(lr){4-5}
        &    	& 	  & \scriptsize{-C} & \scriptsize{(-B\&C)} &&\\
\midrule
        \multirow{4}{*}{{{\centering \LOS}}}
        & Recall@$5$  & 33.4$^*$ & 1.5\textcolor{white}{$^*$} & 6.7$^*$ & 10.5$^*$ & 1.9$^*$\\
        & Recall@$10$ & 28.2$^*$ & -0.6\textcolor{white}{$^*$}& 3.4\textcolor{white}{$^*$} & 9.3$^*$  & 1.5$^*$\\
        & NDCG@$5$    & 34.7$^*$ & 1.6\textcolor{white}{$^*$} & 7.1$^*$ & 11.6$^*$ & 2.1\textcolor{white}{$^*$}\\
        & NDCG@$10$   & 31.3$^*$ & 0.2\textcolor{white}{$^*$} & 4.9\textcolor{white}{$^*$} & 10.4$^*$ & 1.8\textcolor{white}{$^*$}\\
        \bottomrule
      \end{tabular}
      \begin{tablenotes}[normal,flushleft]
      \begin{scriptsize}
      \item
          In this table, 
          the column \Caser/\SASRec/\HGN/\methodMean represents the percentage improvement of \methodCorm over 
          the corresponding method, respectively. 
          %
          For \CUT and \CUS, the column ``\scriptsize{all}"/``\scriptsize{-B}"/``\scriptsize{-C}" 
          represents that the improvement is calculated over all the datasets/with Books excluded/with Comics excluded, 
          respectively. 
	For \LOS, the column ``\scriptsize{-C}"/``\scriptsize{-B\&C}" 
	represents that the improvement is calculated with Comics excluded/with Books and Comics excluded, respectively.
        The $^*$ indicates that the improvement is statistically significant at 90\% confidence level.
          \par
      \end{scriptsize}
      \end{tablenotes}
  \end{threeparttable}
  \vspace{-10pt}
\end{table}


Table~\ref{tbl:impro} shows that overall \methodCorm performs the best, and
achieves significant improvement over other baseline methods in all the 
settings.
For example, compared to \HGN, \methodCorm achieves 12.8\%, 11.9\% 
and 10.5\% improvement in \CUT, \CUS and \LOS, respectively.
This indicates that \methodCorm could achieve the state-of-the-art performance 
regardless of experimental settings.
We notice that in terms of Recall@$10$ and NDCG@$10$, 
\methodCorm achieves very similar performance compared 
to that of \SASRec in \LOS.
This is because \SASRec works very well on the Books dataset.
However, excluding Books, \methodCorm still 
achieves 3.4\% and 4.9\%
improvement over \SASRec on Recall@$10$ and NDCG@$10$, respectively. 
%

Table~\ref{tbl:impro} also shows that \methodCorm achieves highly 
significant improvement over other methods on all the evaluation metrics
in the widely used \CUT setting.
For example, in terms of Recall@$5$, \methodCorm 
achieves 31.6\%, 7.0\%, 12.8\% and 3.1\% improvement over 
\Caser, \SASRec, \HGN and \methodMean in \CUT, respectively.
In {\CUS} and {\LOS}, the improvement of {\methodCorm} over 
\SASRec, \HGN and \methodMean
is slightly lower than that in {\CUT}. 
However, the improvement is still significant.
For example, in terms of Recall@$5$, 
\methodCorm achieves 3.6\%, 11.9\% and 3.2\% 
improvement over \SASRec, \HGN and \methodMean, respectively, in \CUS.
We notice that in \CUS and \LOS, the average improvement of 
\methodCorm over \SASRec is not statistically 
significant on most evaluation metrics. 
This might be due to the superior performance of \SASRec on Books and Comics in \CUS
and on Books in \LOS.
However, \methodCorm still achieves improvement as high as 5.5\% in \CUS  at Recall@$5$ and 7.1\% 
in \LOS at NDCG@$5$.  
Moreover, \methodCorm achieves statistically significant improvement over \SASRec on 
multiple individual datasets (e.g., CDs, Children and ML-20M) 
as shown in Table~\ref{tbl:performance}-\ref{tbl:performancelevelNDCG}. 

%

\subsection{Parameter Study}
\label{sec:results:parameter}


\begin{table}[!h]
\footnotesize
  \caption{\mbox{Parameter Study of \methodCorm on CDs in \CUT}}
    \label{tbl:paraCDs}
  \centering
  \begin{threeparttable}
      \begin{tabular}{
        @{\hspace{0pt}}c@{\hspace{1pt}}
        @{\hspace{1pt}}r@{\hspace{8pt}}
        @{\hspace{8pt}}r@{\hspace{8pt}}
        @{\hspace{8pt}}r@{\hspace{8pt}}
        @{\hspace{8pt}}r@{\hspace{8pt}}
        @{\hspace{10pt}}r@{\hspace{8pt}}
        @{\hspace{6pt}}r@{\hspace{8pt}}
    @{\hspace{2pt}}r@{\hspace{0pt}}
        }
        \toprule
        parameter & $d$ & $\LL$ & $\M$ & $\T$ & $p$ & Recall@5 & Recall@10 \\
        \midrule
        \multirow{4}{*}{{{\centering $d$}}}
        & 200 & 5 & 2 & 3 & 2 & 0.0392 & 0.0603\\
        & \underline{400} & \underline{5} & \underline{2} & \underline{3} & \underline{2} 
        & \underline{0.0397} & \underline{0.0615}\\
        & 600 & 5 & 2 & 3 & 2 & \textbf{$\mathclap{^{\dagger~}}$0.0403} & \textbf{0.0618}\\
        & 800 & 5 & 2 & 3 & 2 & 0.0400 & 0.0613\\
        \midrule
        \multirow{4}{*}{{{\centering $\LL$}}}
        & 400 & 3 & 2 & 3 & 2 & 0.0391 & 0.0610\\
        & 400 & 4 & 2 & 3 & 2 & \textbf{0.0397} & \textbf{0.0617}\\
        & \underline{400} & \underline{5} & \underline{2} & \underline{3} & \underline{2} 
        & \underline{\textbf{0.0397}} & \underline{0.0615}\\
        & 400 & 6 & 2 & 3 & 2 & 0.0378 & 0.0608\\
        \midrule
        \multirow{4}{*}{{{\centering $\M$}}}
        & 400 & 5 & 0 & 3 & 2 & 0.0378 & 0.0597\\
        & 400 & 5 & 1 & 3 & 2 & \textbf{0.0398} & 0.0606\\
        & \underline{400} & \underline{5} & \underline{2} & \underline{3} & \underline{2} 
        & \underline{0.0397} & \underline{\textbf{0.0615}}\\
        & 400 & 5 & 3 & 3 & 2 & 0.0394 & 0.0607\\
        \midrule
        \multirow{3}{*}{{{\centering $\T$}}}
        & 400 & 5 & 2 & 2 & 2 & 0.0380 & 0.0604\\
        & \underline{400} & \underline{5} & \underline{2} & \underline{3} & \underline{2} 
        & \underline{\textbf{0.0397}} & \underline{\textbf{0.0615}}\\
        & 400 & 5 & 2 & 4 & 2 & 0.0389 & 0.0603\\
        \midrule
        \multirow{3}{*}{{{\centering $p$}}}
        & 400 & 5 & 2 & 3 & 1 & \textbf{$\mathclap{^{\dagger~}}$0.0403} & \textbf{$\mathclap{^{\dagger~}}$0.0625}\\ 
        & \underline{400} & \underline{5} & \underline{2} & \underline{3} & \underline{2} 
        & \underline{0.0397} & \underline{0.0615}\\
        & 400 & 5 & 2 & 3 & 3 & 0.0378 & 0.0596\\
        \bottomrule
      \end{tabular}
      \begin{tablenotes}[normal,flushleft]
      \begin{scriptsize}
      \item
          In this table, $d$, $\LL$/$\M$, $\T$ and $p$ are
          the dimension of embeddings, number of items in high-order/low-order associations,
          number of items to calculate recommendation errors during training
          and the order of item synergies.
	  The best performance overall is \textbf{~$\mathclap{^{\dagger~}}$bold}
          and the best performance in each row block is \textbf{bold}.
          The best results based on validation sets and the corresponding parameters tuned on the validation sets are \underline{underlined}.
          The ``parameter'' column presents the parameter to be studied in each row block.
          \par
      \end{scriptsize}
      \end{tablenotes}
  \end{threeparttable}
\end{table}


\begin{table}[!h]
\footnotesize
  \caption{\mbox{Parameter Study of \methodCorm on Children in \CUT}}
    \label{tbl:paraChild}
  \centering
  \begin{threeparttable}
      \begin{tabular}{
        @{\hspace{0pt}}c@{\hspace{1pt}}
        @{\hspace{1pt}}r@{\hspace{8pt}}
        @{\hspace{8pt}}r@{\hspace{8pt}}
        @{\hspace{8pt}}r@{\hspace{8pt}}
        @{\hspace{8pt}}r@{\hspace{8pt}}
        @{\hspace{10pt}}r@{\hspace{8pt}}
        @{\hspace{6pt}}r@{\hspace{8pt}}
    @{\hspace{2pt}}r@{\hspace{0pt}}
        }
        \toprule
        parameter & $d$ & $\LL$ & $\M$ & $\T$ & $p$ & Recall@5 & Recall@10 \\
        \midrule
        \multirow{3}{*}{{{\centering $d$}}}
        & 200 & 6 & 1 & 4 & 3 & 0.0917 & \textbf{0.1395}\\
        & \underline{400} & \underline{6} & \underline{1} & \underline{4} & \underline{3} 
        & \underline{\textbf{$\mathclap{^{\dagger~}}$0.0921}} & \underline{0.1393}\\
        & 600 & 6 & 1 & 4 & 3 & 0.0908 & 0.1372\\
        \midrule
        \multirow{3}{*}{{{\centering $\LL$}}}
        & 400 & 5 & 1 & 4 & 3 & 0.0917 & 0.1390\\
        & \underline{400} & \underline{6} & \underline{1} & \underline{4} & \underline{3} 
        & \underline{\textbf{$\mathclap{^{\dagger~}}$0.0921}} & \underline{\textbf{0.1393}}\\
        & 400 & 7 & 1 & 4 & 3 & 0.0909 & 0.1383\\
        \midrule
        \multirow{3}{*}{{{\centering $\M$}}}
        & \underline{400} & \underline{6} & \underline{1} & \underline{4} & \underline{3}
        & \underline{\textbf{$\mathclap{^{\dagger~}}$0.0921}} & \underline{\textbf{0.1393}}\\
        & 400 & 6 & 2 & 4 & 3 & 0.0897 & 0.1370\\
        & 400 & 6 & 3 & 4 & 3 & 0.0875 & 0.1339\\
        \midrule
        \multirow{3}{*}{{{\centering $\T$}}}
        & 400 & 6 & 1 & 3 & 3 & 0.0906 & 0.1371\\
        & \underline{400} & \underline{6} & \underline{1} & \underline{4} & \underline{3}
        & \underline{\textbf{$\mathclap{^{\dagger~}}$0.0921}} & \underline{\textbf{0.1393}}\\
        & 400 & 6 & 1 & 5 & 3 & 0.0914 & 0.1391\\
        \midrule
        \multirow{3}{*}{{{\centering $p$}}}
        & 400 & 6 & 1 & 4 & 1 & 0.0904 & 0.1383\\
        & 400 & 6 & 1 & 4 & 2 & 0.0914 & \textbf{$\mathclap{^{\dagger~}}$0.1396}\\
        & \underline{400} & \underline{6} & \underline{1} & \underline{4} & \underline{3}
        & \underline{\textbf{$\mathclap{^{\dagger~}}$0.0921}} & \underline{0.1393}\\
        & 400 & 6 & 1 & 4 & 4 & 0.0903 & 0.1374\\
        \bottomrule
      \end{tabular}
      \begin{tablenotes}[normal,flushleft]
      \begin{scriptsize}
      \item
      The columns in this table have the same meanings as those in Table~\ref{tbl:paraCDs}. 
          \par
      \end{scriptsize}
      \end{tablenotes}
  \end{threeparttable}
  \vspace{-15pt}
\end{table}


We conduct a comprehensive parameter study for \methodCorm on the 
most sparse dataset CDs and moderately sparse datasets Children and Comics
in the widely used \CUT setting, following \HGN.
%
%
In the study, we first identify the parameter values that achieve the best performance
on the validation sets of CDs, Children and Comics.
Then, we change one of the values and fix the others so as to study 
how the changing parameter affects the recommendation performance. 
We report the corresponding results on the testing sets of CDs and Children
in Table~\ref{tbl:paraCDs} and Table~\ref{tbl:paraChild}, and the results of Comics
in Table~\ref{tbl:paraComics}, respectively. 
The best results based on validation sets (Table~\ref{tbl:performance})  and the corresponding parameters 
tuned on the validation sets are {underlined} in Table~\ref{tbl:paraCDs}, 
Table~\ref{tbl:paraChild} and Table~\ref{tbl:paraComics}. 
%
%
Please note that the {underlined} parameters are the best parameters based 
on the tuning on the validation set but they are not necessarily the best parameters 
on the testing set. 
Thus, other parameters may achieve even 
better performance on the testing set. We use different values on the testing set just for parameter study purposes, but 
using the parameter values tuned on the validation set is the proper way to compare different methods 
as in Section~\ref{sec:experiments:setting:cut}. 
Overall, as shown in Table~\ref{tbl:paraCDs}, \ref{tbl:paraChild} and \ref{tbl:paraComics}, 
\methodCorm is stable with respect to the parameters within a certain optimal range. 
%
In addition, \methodCorm is able to achieve the state-of-the-art performance on different datasets 
with appropriate parameters.



%
\subsubsection{Parameter Study of \methodCorm on CDs in \CUT}
\label{sec:results:parameter:cds}
%
Table~\ref{tbl:paraCDs} presents the results of parameter study on CDs. 
As shown in this table, 
\methodCorm achieves better performance as the dimension of embeddings $d$ increases, 
and achieves the best performance with $d$$=$$600$. 
%
This indicates that \methodCorm is able to learn from 
even sparse datasets into large-dimension embeddings. 
%
%
%
Table~\ref{tbl:paraCDs} also shows that \methodCorm achieves the best performance when 
association orders are small 
($\LL$$=$$4$, $\M$$=$$2$).
This indicates that the most recent associations 
among a few items are effective in inducing the next items. 
It also demonstrates the importance of modeling low-order associations in \method.
Table~\ref{tbl:paraCDs} also shows that
as more items are used to calculate recommendation errors during training 
(i.e., larger {$\T$}, Fig.~\ref{fig:setting}), 
the performance increases first and then decreases, and 
\methodCorm achieves the best performance with $\T$$=$$3$.
%
It indicates that the association patterns are most effective in 
inducing the next few items than the items purchased/rated much later.

We also notice that {\methodCorm} has poor performance as the order of 
item synergies $p$ is large.
This conforms to the results in Table~\ref{tbl:performance}
that \methodCorm underperforms \methodMean (i.e., \methodCorm without item synergies) on CDs in \CUT.
This might be due to the same reason as discussed in Section~{\ref{sec:results:lof:chh}},  
that is, the CDs dataset is extremely sparse and thus, very limited information about item synergies 
can be learned from the data.
However,
with $p$$=$$2$ (Recall@$5$$=$$0.0397$), the performance of \methodCorm is still very 
comparable to that without item synergies (i.e., $p$$=$$1$, Recall@$5$=0.0403, difference 1.5\%), 
indicating that \methodCorm could still achieve the state-of-the-art performance even on 
extremely sparse dataset.
%
Moreover, as will be shown later in 
Table~\ref{tbl:paraChild}
and Table~\ref{tbl:paraComics}, 
\methodCorm could achieve highly significant 
improvement after incorporating item synergies on Children and Comics, 
indicating that explicitly model item synergies will 
improve the recommendation performance on most benchmark datasets.
Please note that we tried $p$ up to $\LL$ in the study
but did not present the results 
when unnecessary given the clear trend.


\subsubsection{Parameter Study of \methodCorm on Children in \CUT}
\label{sec:results:parameter:children}
%
Table~\ref{tbl:paraChild} presents the results of parameter study on Children.
Similar to that on CDs, on Children, 
\methodCorm  achieves the best performance 
with a large $d$ (i.e., 400) and a small $\M$ (i.e., 1).
However, different with that on CDs, 
\methodCorm achieves the best performance on Children
with relatively large $\LL$ and $\T$ ($\LL$$=$$6$, $\T$$=$$4$). 
This is probably due to that there could be a lot of associations with high orders on Children.
%
In addition, \methodCorm achieves the best performance with large $p$ (i.e., 3).
Compared to \methodCorm without item synergies (i.e., $p$$=$$1$), 
\methodCorm with item synergies achieves reasonable improvement 
(e.g., $p$$=$$3$: 0.0921 vs  $p$$=$$1$: 0.0904 on Recall@$5$, 1.9\% improvement), 
demonstrating that item synergies could effectively improve the recommendation performance.
%

%

%

%
%
%


\subsubsection{Parameter Study of \methodCorm on Comics in \CUT}
\label{sec:results:parameter:comics}

\begin{table}[!h]
\footnotesize
  \caption{\mbox{Parameter Study of \methodCorm on Comics in \CUT}}
  \label{tbl:paraComics}
  \centering
  \begin{threeparttable}
      \begin{tabular}{
        @{\hspace{0pt}}c@{\hspace{1pt}}
        @{\hspace{1pt}}r@{\hspace{8pt}}
        @{\hspace{8pt}}r@{\hspace{8pt}}
        @{\hspace{8pt}}r@{\hspace{8pt}}
        @{\hspace{8pt}}r@{\hspace{8pt}}
        @{\hspace{10pt}}r@{\hspace{8pt}}
        @{\hspace{6pt}}r@{\hspace{8pt}}
        @{\hspace{2pt}}r@{\hspace{0pt}}
        }
        \toprule
        parameter & $d$ & $\LL$ & $\M$ & $\T$ & $p$ & Recall@5 & Recall@10 \\
        \midrule
        \multirow{3}{*}{{{\centering $d$}}}
        & 200 & 7 & 2 & 5 & 3 & 0.1354 & 0.1914\\
        & \underline{400} & \underline{7} & \underline{2} & \underline{5} & \underline{3} 
        & \underline{\textbf{0.1385}} & \underline{\textbf{0.1945}}\\
        & 600 & 7 & 2 & 5 & 3 & 0.1378 & 0.1929\\
        \midrule
        \multirow{4}{*}{{{\centering $\LL$}}}
        & 400 & 6 & 2 & 5 & 3 & 0.1372 & 0.1930\\
        & \underline{400} & \underline{7} & \underline{2} & \underline{5} & \underline{3} 
        & \underline{0.1385} & \underline{0.1945}\\
        & 400 & 8 & 2 & 5 & 3 & \textbf{0.1389} & 0.1945\\
        & 400 & 9 & 2 & 5 & 3 & 0.1387 & \textbf{0.1948}\\
        \midrule
        \multirow{4}{*}{{{\centering $\M$}}}
        & 400 & 7 & 0 & 5 & 3 & 0.1223 & 0.1767\\
        & 400 & 7 & 1 & 5 & 3 & \textbf{$\mathclap{^{\dagger~}}$0.1398} & \textbf{$\mathclap{^{\dagger~}}$0.1953}\\
        & \underline{400} & \underline{7} & \underline{2} & \underline{5} & \underline{3} 
        & \underline{0.1385} & \underline{0.1945}\\
        & 400 & 7 & 3 & 5 & 3 & 0.1352 & 0.1902\\
        \midrule
        \multirow{3}{*}{{{\centering $\T$}}}
        & 400 & 7 & 2 & 4 & 3 & 0.1380 & 0.1930\\
        & \underline{400} & \underline{7} & \underline{2} & \underline{5} & \underline{3} 
        & \underline{\textbf{0.1385}} & \underline{\textbf{0.1945}}\\
        & 400 & 7 & 2 & 6 & 3 & 0.1367 & 0.1944\\
        \midrule
        \multirow{3}{*}{{{\centering $p$}}}
        & 400 & 7 & 2 & 5 & 1 & 0.1299 & 0.1874\\
        & 400 & 7 & 2 & 5 & 2 & 0.1331 & 0.1888\\
        & \underline{400} & \underline{7} & \underline{2} & \underline{5} & \underline{3} 
        & \underline{\textbf{0.1385}} & \underline{\textbf{0.1945}}\\
        & 400 & 7 & 2 & 5 & 4 & 0.1311 & 0.1859\\
        \bottomrule
      \end{tabular}
      \begin{tablenotes}[normal,flushleft]
      \begin{scriptsize}
      \item
          The columns in this table have the same meanings as those in Table~\ref{tbl:paraCDs}.
          \par
      \end{scriptsize}
      \end{tablenotes}
  \end{threeparttable}
\end{table}



Table~\ref{tbl:paraComics} presents the results of parameter study on the Comics dataset.
As shown in this table, similar to that on CDs and Children, on Comics, 
\methodCorm still achieves the best performance with a large $d$ (i.e., 400)
and a small $\M$ (i.e., 2).
Similar to that on Children, \methodCorm achieves the best performance 
with relatively large $\LL$, $\T$ ($\LL$$=$$8$, $\T$$=$$5$) and $p$ (i.e, 3).
On Comics, compared to \methodCorm without item synergies (i.e., $p$$=$$1$), 
\methodCorm with item synergies achieves significant improvement
(e.g., $p$$=$$3$: 0.1385 vs  $p$$=$$1$: 0.1299 on Recall@$5$, 6.6\% improvement),
demonstrating that item synergies could improve the recommendation performance 
on most benchmark datasets.
We also notice that as shown in Table~\ref{tbl:paraChild}
and Table~\ref{tbl:paraComics}
on Children and Comics,
the best parameters based on validation sets
achieve the best or near the best performance on testing as well.

\subsection{Ablation Study}
\label{sec:results:abla}

\begin{table}[!t]
\footnotesize
  \caption{\mbox{Ablation Study of \methodCorm in \CUT}}
    \label{tbl:abla}
  \centering
  \begin{threeparttable}
      \begin{tabular}{
        @{\hspace{1pt}}l@{\hspace{3pt}}
        @{\hspace{3pt}}l@{\hspace{6pt}}
        @{\hspace{6pt}}r@{\hspace{6pt}}
        @{\hspace{6pt}}r@{\hspace{6pt}}
        @{\hspace{6pt}}r@{\hspace{6pt}}
        @{\hspace{6pt}}r@{\hspace{6pt}}
        @{\hspace{6pt}}r@{\hspace{6pt}}
        @{\hspace{3pt}}r@{\hspace{3pt}}
    @{\hspace{3pt}}r@{\hspace{3pt}}
        }
        \toprule
        Dataset & model & $d$ & $\LL$ & $\M$ & $\T$ & $p$ & Recall@5 & Recall@10 \\
        \midrule
        \multirow{3}{*}{CDs} 
        & \methodCorm & 400 & 5 & 2 & 3 & 2 & 0.0397 & 0.0615\\
        & \methodCormo
        & 400 & 3 & 0 & 3 & 2 & \textbf{0.0403} 
        & \textbf{0.0623}\\
        & \methodCormu
        & 400 & 6 & 2 & 2 & 2 & 0.0346 & 0.0549\\
        \midrule
        \multirow{3}{*}{Books}
        & \methodCorm & 400 & 9 & 2 & 7 & 2 & \textbf{0.0412}
        & \textbf{0.0630}\\
        & \methodCormo & 400 & 3 & 0 & 8 & 2 & 0.0367 & 0.0579\\
        & \methodCormu & 400 & 7 & 1 & 5 & 2 & 0.0373 & 0.0548\\
        \midrule
        \multirow{3}{*}{Children}
        & \methodCorm & 400 & 6 & 1 & 4 & 3 & \textbf{0.0921}
        & \textbf{0.1393}\\
        & \methodCormo & 600 & 3 & 0 & 4 & 3 & 0.0854 & 0.1324\\
        & \methodCormu & 200 & 6 & 1 & 4 & 2 & 0.0910 & 0.1386\\
        \midrule
        \multirow{3}{*}{Comics}
        & \methodCorm & 400 & 7 & 2 & 5 & 3 & 0.1385 & 0.1945\\
        & \methodCormo & 400 & 6 & 0 & 3 & 3
        & 0.1205 & 0.1726\\
        & \methodCormu & 200 & 7 & 1 & 6 & 2
        & \textbf{0.1422} 
        & \textbf{0.1974}\\
        \midrule
        \multirow{3}{*}{ML-20M}
        & \methodCorm     & 400 & 9  & 3 & 2 & 3 & \textbf{0.0838}
        & 0.1389\\
        & \methodCormo  & 600 & 8  & 0 & 3 & 3 & 0.0834 & \textbf{0.1394}\\
        & \methodCormu & 600 & 10 & 3 & 3 & 2 & 0.0830 & 0.1379\\
        \midrule
        \multirow{3}{*}{ML-1M}
        & \methodCorm & 400 & 7 & 2 & 3 & 3 & \textbf{0.0793}
        & \textbf{0.1330}\\
        & \methodCormo & 600 & 5 & 0 & 1 & 3 & 0.0763 & 0.1291\\
        & \methodCormu & 200 & 9 & 2 & 3 & 2 & 0.0792 & 0.1322\\
        \bottomrule
      \end{tabular}
      \begin{tablenotes}[normal,flushleft]
      \begin{scriptsize}
      \item
          In this table, $d$, $\LL$/$\M$, $\T$ and $p$ are 
	  the dimension of embeddings, number of items in high-order/low-order associations,
          number of items to calculate recommendation errors during training
          and the order of item synergies.
	  The best performance on each dataset is \textbf{bold}.
          The ``model'' column presents the factor ablated:
          the row corresponding to \methodCorm represents the full model without any ablation; 
          the row corresponding to \methodCormo 
          represents the results when the low-order association is ablated (i.e., $\M$$=$$0$);  
 	  the row corresponding to \methodCormu represents the results when the users' general preferences is ablated. 
          \par
      \end{scriptsize}
      \end{tablenotes}
  \end{threeparttable}
  \vspace{-10pt}
\end{table}



We also conduct an ablation study for \methodCorm 
in \CUT. 
In particular, we want to verify the effect 
of the low-order associations $\sse$ (i.e., only one item association term) 
and users' general preferences $\user$
in \method methods from the ablation study.
To this end, we remove $\sse$ or $\user$ from the objective 
function (Equation~\ref{eqn:obj}) and tune parameters for the ablated \methodCorm
on the validation sets.
We report the results on the testing sets 
in Table~\ref{tbl:abla}, where we denote the \methodCorm without $\sse$/$\user$
as \methodCormo/\methodCormu, respectively.

Table~\ref{tbl:abla} shows that overall \methodCorm achieves the best 
performance compared to \methodCormo and \methodCormu.
In particular, \methodCorm achieves the best performance on 4 out of 6 datasets
and the second best performance on the rest CDs and Comics datasets.
Table~\ref{tbl:abla} also shows that \methodCorm significantly outperforms \methodCormo 
(i.e., \methodCorm without $\sse$)
on the 4 datasets.
For example, in terms of Recall@$5$, \methodCorm significantly 
outperforms 
\methodCormo by 12.3\%, 7.8\%, 14.9\% and 3.9\% on 
Books, Children, Comics and ML-1M, respectively.
This indicates that modeling both high and low-order associations could significantly 
improve the recommendation performance.
We also notice that \methodCorm achieves very similar performance 
with \methodCormo on CDs and ML-20M.
(e.g., \methodCorm: 0.0397 vs \methodCormo: 0.0403 on Recall@$5$ on CDs, 
difference: 1.5\%). 
On CDs, this might be due to that \methodCormo uses 
a small $\LL$$=$$3$ to model the item associations, and it would have a similar effect 
as to \methodCorm
using high-order $\LL$$=$$5$ and low-order $\M$$=$$2$ together.
On ML-20M, this might be because of ML-20M 
has very limited low-order associations.
Thus, removing $\sse$ doesn't significantly 
affect the recommendation performance.
%

Table~\ref{tbl:abla} shows that \methodCorm significantly outperforms 
\methodCormu (i.e., \methodCorm without $\user$) on 5 out of 6 datasets 
(i.e., CDs, Books, Children, ML-20M and ML-1M).
This indicates the effectiveness of explicitly modeling users' general preferences.
We also notice that \methodCorm is slightly worse  
than \methodCormu on Comics 
(e.g., \methodCorm: 0.1385 vs \methodCormu: 0.1422 on Recall@$5$, 
difference: 2.6\%).
This indicates that users may have limited long-term preferences on comics.
Thus, explicitly modeling the long-term preferences 
does not improve the recommendation performance.
%
Overall, \methodCorm significantly outperforms \methodCormo and \methodCormu on most 
benchmark datasets, demonstrating the effectiveness of explicitly modeling 
high-order and low-order associations and users' general preferences.

\subsection{Run-Time Performance in Testing}
\label{sec:results:time}
\begin{table}[t!]
\footnotesize
  \captionsetup{justification=centering}
  \caption{\mbox{{Testing Run-Time Performance in {\CUT} (sec)}}}
  \centering
  \label{tbl:time}
  \begin{threeparttable}
      \begin{tabular}{
        @{\hspace{12pt}}l@{\hspace{8pt}}
        @{\hspace{8pt}}r@{\hspace{8pt}}
        @{\hspace{4pt}}r@{\hspace{8pt}}
        @{\hspace{12pt}}r@{\hspace{8pt}}
        @{\hspace{12pt}}r@{\hspace{4pt}}
        @{\hspace{8pt}}c@{\hspace{12pt}}
        }
        \toprule
         Dataset & \Caser & \SASRec & \HGN & \methodCorm & speedup\\
        \midrule
         CDs      & 1.2e-1 & 2.3e-2 & \underline{1.5e-3} & \textbf{6.3e-4} & 2.4\\
         Books    & 1.4e-1 & 2.6e-2 & \underline{3.2e-3} & \textbf{7.9e-4} & 4.1\\
         Children & 1.2e-1 & 1.9e-2 & \underline{1.6e-3} & \textbf{7.4e-4} & 2.2\\
         Comics   & 1.2e-1 & 2.2e-2 & \underline{1.5e-3} & \textbf{7.1e-4} & 2.1\\
         ML-20M   & 5.0e-2 & 1.2e-2 & \underline{6.0e-4} & \textbf{4.9e-4} & 1.2\\
         ML-1M    & 1.3e-2 & 6.1e-3 & \underline{3.7e-4} & \textbf{3.5e-4} & 1.1\\
        \bottomrule
      \end{tabular}
      \begin{tablenotes}[normal,flushleft]
      \begin{scriptsize}
      \item
          In this table, the best run-time performance in testing of each dateset is \textbf{bold}.
          The second best inference-time performance in each dataset is \underline{underlined}.
          All the run time is in seconds (s). 
          The column ``speedup'' presents the speedup of the \textbf{bold} best performance
          over the \underline{underlined} second best performance in each dataset.
          \par
      \end{scriptsize}
      \end{tablenotes}
  \end{threeparttable}
  \vspace{-10pt}
\end{table}


Table~\ref{tbl:time} presents the run-time performance during testing 
for each user on average (i.e., on the testing set)
in \CUT.
We compare the run-time performance in testing instead of training 
due to the fact that in real applications, latency in real-time recommendation affects the user 
experience and thus revenue very significantly, while model training can be implemented as an off-line or online, 
incremental/batch process and thus long run time for model training can be tolerated. 
%
%

%
Table~\ref{tbl:time} shows that \methodCorm achieves the best run-time performance in testing
compared to all the baseline methods on all the datasets, followed by \HGN. 
\methodCorm is on average 2.2-times faster than the best baseline \HGN over all the datasets. 
The primary difference between \HGN and \methodCorm during testing is that
\HGN needs to calculate weights for latent features and items 
using the gating mechanism, but \methodCorm only uses mean pooling to 
aggregate items.
%
%
%
Compared to deep methods \SASRec and \Caser, \methodCorm achieves highly 
significant speedup on all the six datasets.
For example, \methodCorm achieves 190.5-times 
(e.g, 1.2e-1/6.3e-4 = 190.5) faster than \Caser
on CDs.
On average, \methodCorm achieves 139.7 and 28.0 speedup 
compared to \Caser and \SASRec, respectively.
%
This indicates that by using the simplistic pooling mechanism, \methodCorm could be 
much more efficient than deep methods such as \SASRec and \Caser.
Recall that as shown in Table~\ref{tbl:impro} (Section~\ref{sec:results:all}), 
in \CUT, \methodCorm also achieves significant improvement over \Caser, \SASRec  and \HGN
in terms of the recommendation performance. 
These results indicate that \methodCorm  is both much more effective 
and much more efficient than state-of-the-art methods.
We also notice that 
although \methodCorm is slightly worse than \Caser and \SASRec on the ML-1M and Books datasets, 
respectively, in terms of the recommendation performance 
(Table~\ref{tbl:performance} and Table~\ref{tbl:performanceNDCG}), 
\methodCorm is much faster than \Caser and \SASRec: 
\methodCorm achieves 37.1-fold (1.3e-2/3.5e-4$=$37.1) and 32.9-fold (2.6e-2/7.9e-4$=$32.9) speedup over 
\Caser on ML-1M and \SASRec on Books, respectively.
%
%
Overall, \methodCorm significantly outperforms state-of-the-art baseline methods 
in run-time performance in testing, demonstrating its efficiency.

Similar trend is also observed in the run-time performance during training. 
For example, in \CUT, on Books, \methodCorm needs 1.1 hour to achieve the reported results 
(Table~\ref{tbl:performance}), while \Caser and \SASRec needs 3.5 hours (3-fold slower) 
and 7.7 hours (7-fold slower), respectively. 
\methodCorm could be slower than \HGN in training due to the fact that \methodCorm 
generally needs more epochs to converge than \HGN, while \HGN has lower embedding dimensions 
(Table~\ref{tbl:para:ham} in Appendix).
However, we believe the difference is very tolerable in real applications.
For example, in \CUT, on Books, \methodCorm needs 1.1 hours to achieve the reported results, while 
\HGN needs 0.7 hours (24-minute difference).
On ML-20M (i.e., the largest dataset), the difference is relatively large:  
\methodCorm takes 4.0 hours to achieve the reported best results, while 
\HGN needs 1.0 hours (3-hour difference).
However, we think the 3-hours difference in training is tolerable in real applications, particularly given 
that the training can be done off line.
%

%

%

\section{Discussions}
\label{sec:discussion}

\subsection{Comparison with Reported Results}
\label{sec:discussion:para}


In \HGN~\cite{ma2019hierarchical}, the authors reported that \HGN significantly outperforms 
\SASRec on the CDs, Books, Children, Comics and ML-20M datasets in \CUT.
For example,  \HGN achieves 19.83\% 
and 16.67\%~\cite{ma2019hierarchical} improvement over \SASRec 
on Books and Comics, respectively, in terms of Recall@$10$ in \CUT.
However, in our experiments, as shown in Table~\ref{tbl:performance}, 
\SASRec outperforms \HGN on 4 out of 6 datasets (i.e., CDs, Books, ML-20M and ML-1M), 
%
%
after we used a much larger parameter space and exhaustively tuned more parameters (e.g., $d$, $\LL$, $\T$) 
than those reported in \HGN.
%
However, as we have discussed in Section~\ref{sec:results:loo}, 
\SASRec is sensitive to parameters and it becomes very expensive
to find the best performance of \SASRec in real applications. 
This could be the reason why we observed the discordance 
between our experimental results and 
those reported in \HGN.

%

\subsection{\HGN Attention Weight Analysis}
\label{sec:discussion:weight}

\begin{figure}[!t]
  \vspace{-10pt}
  \centering
  \input{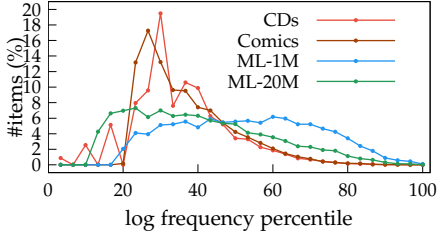}
  \vspace{-5pt}
  \captionof{figure}{Item Frequency Distribution}
  \label{fig:total_distribution}
  \vspace{-15pt}
\end{figure}
%

It has been shown that the learned attention weights may not always be 
meaningful~\cite{jain2019attention,serrano2019attention}.
Therefore, we further investigate the attention weights in \HGN to 
interpret their significance and to understand why instead the simplistic mean 
pooling in \method would suffice. 
We use datasets CDs, Comics, ML-1M and ML-10M in the investigation, 
because these datasets, as Table~\ref{tbl:dataset} shows, 
represent different data sparsities (i.e., CDs is highly sparse, 
Comics is moderately sparse, and ML-1M and ML-10M are dense).  
Fig.~\ref{fig:total_distribution} (the x-axis in the Fig. is logarithmized item frequencies 
and then normalized into
[0,1]) shows that most of the items in CDs and Comics are very 
infrequent, 
whereas in ML-1M infrequent items 
are fewer; in ML-20M, the infrequent items (compared to other frequent items in ML-20M) still have 
many purchases/ratings (Table~\ref{tbl:dataset}). 

\begin{figure}[!t]
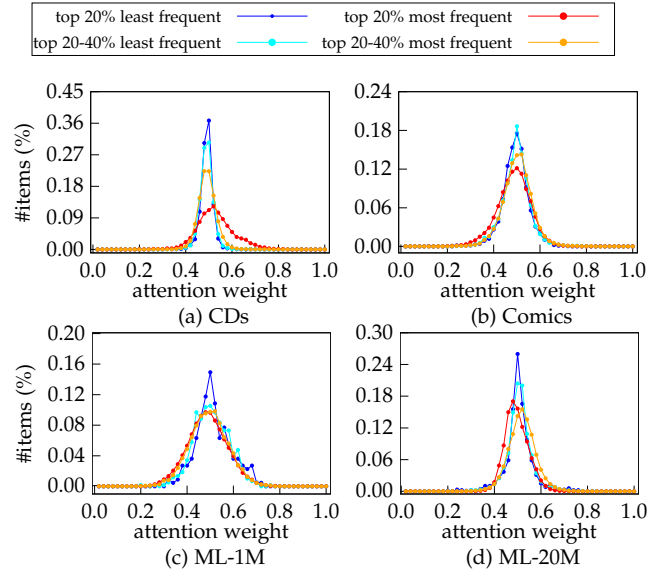

  \centering
  \begin{minipage}{\linewidth}
    \begin{center}
    \scalebox{0.9}{
      {\input{plots/CDs/legend.tex}}
      }
    \end{center}
  \end{minipage}
%

  \begin{minipage}{\linewidth}
  \hspace{15pt}
	\centering
	\begin{subfigure}{0.45\linewidth}
		\centering
		\input{./plots/CDs/CDs.tex}	
		\vspace{13pt}
		\caption{CDs}
		\label{fig:cd_weights}
	\end{subfigure}
	\begin{subfigure}{0.45\linewidth}
                \centering
                \input{./plots/Comics/Comics.tex}
                \vspace{13pt}
                \caption{Comics}
                \label{fig:comics_weights}

	\end{subfigure}
       \end{minipage}

  \begin{minipage}{\linewidth}

 \hspace{18pt}
	\begin{subfigure}{0.45\linewidth}


                \centering
                \input{./plots/ml1m/ml1m.tex}
                \vspace{13pt}
                \caption{ML-1M}
                \label{fig:ml1m_weights}
	\end{subfigure}
        \begin{subfigure}{0.45\linewidth}
                \centering
                \input{./plots/ml20m/ml20m.tex}
                \vspace{13pt}
                \caption{ML-20M}
                \label{fig:ml20m_weights}
        \end{subfigure}
\end{minipage}
\caption{\HGN Attention Weight Distributions}
\label{fig:datasets_plots}
\vspace{-15pt}
\end{figure}
%


Fig.~\ref{fig:cd_weights}, \ref{fig:comics_weights}, \ref{fig:ml1m_weights}
and \ref{fig:ml20m_weights} present the 
distributions of attention weights from the best performing \HGN models on the CDs, 
Comics, ML-1M and ML-20M, respectively.  
Note that a same item can have different weights 
in different user sequences; we use all the weights of a same item from all the users in the Fig.s.
The distributions for CDs, Comics and ML-20M datasets show very similar pattens: for very infrequent
items, their attention weights 
are highly centered around 0.5 (i.e., the initialization value); 
for the most frequent items, their attention weights are slightly 
off 0.5 and have some different values than 0.5.  
Given that as Fig.~\ref{fig:total_distribution} shows, most of the items in CDs, Comics and ML-20M
are infrequent, the weight distribution over infrequent items indicates that the weights might not be 
well learned to differentiate the importance of infrequent items. The weights on frequent items 
might be relatively better learned. 
However, unfortunately, frequent items are not many and their weights may not 
substantially affect recommendations. 
The distributions for the ML-1M dataset also shows similar pattens: the weights for both 
infrequent and frequent items are closely centered at 0.5, indicating that such weights might not well 
differentiate item importance. 
Such \HGN weight distributions from both sparse and dense datasets indicate that the learned weights
may not play an effective role in recommendation. Instead, a special case of weights, that is, equal weights
as we have in \method, should also achieve comparable performance as \HGN. 
As a matter of fact, equal weights on better learned item representations as in \method actually 
improve the recommendation performance. 


\subsection{Discussions on Experimental Settings}
\label{sec:discussion:settings}


\CUT (i.e., to evaluate the last 20\% items) and NDCG@$k$
(e.g., $k$=10)
are the most commonly used experimental setting and evaluation metric in literature. 
NDCG@$k$ in \CUT could over estimate the sequential recommendation performance
as discussed in Section~\ref{sec:results:lot:complof},  
particularly for long sequences in which the last 20\% items include many. In such long sequences, 
NDCG@$k$ could be high when items purchased/rated very late
are recommended on top. However, such recommendations would have limited use scenarios. 
Meanwhile, many testing items will increase the chances that the recommended $k$ items will be included
in those items, and thus inflate the NDCG@$k$ values. 

\CUS and \LOS mitigate the over-estimation issue because always a same number of items will 
be tested and NDCG@$k$ calculated over a same number of testing items will not be affected by the 
number of testing items.  
However, in \CUS, the number of validation items 
are larger than that of testing items for most users.
As a result, 
the best parameters selected based on validation sets may not 
always work well in recommending the next few items on the testing set.
\LOS mitigates this problem to some extent by using the same number of items (i.e., 3) in both the validation 
and testing sets.
However, there are only three validation items for each user.
The small number of validation items may result in more randomness in identifying the best parameters  
and thus the best parameters on validation sets may not work well on testing either, 
especially for methods that are sensitive to parameters such as 
\SASRec (Section~\ref{sec:results:loo}).

Overall, considering the advantages and disadvantages of the three settings, we think \CUS is more suitable 
for the sequential recommendation problem compared to \CUT and \LOS. 
However, it is still an open question as to how to standardize the experimental settings and evaluation metrics for sequential 
recommendation tasks. 
%
%
%

\section{Conclusions}
\label{sec:conclusions}

In this manuscript, we present novel, effective 
and efficient \method models for sequential recommendation. 
The \method models use three factors to generate recommendations: 
1) users’ long-term preferences, 
2) sequential, high-order and low-order association patterns 
in the users’ most recent purchases/ratings, 
and 3) synergies among those items.
Our experimental results in comparison with three state-of-the-art sequential 
recommendation methods on six public benchmark datasets in three experimental settings 
demonstrate that \method models achieve significant improvement over 
the state-of-the-art baseline methods 
(e.g., 46.6-fold improvement over \Caser in \CUS).
In addition, our experimental results in the ablation study demonstrate the
importance of modeling both low-order associations and users' general preferences
(e.g., 14.9\% improvement with low-order associations modeled on Comics). 
Moreover, our results in the {run-time} performance comparison in testing show 
that \method models are much more efficient than baseline methods especially for 
deep methods such as \Caser (139.7 average speedup) and \SASRec (28.0 average speedup).
%
%
In addition, we studied the attention weights in benchmark datasets and 
found that the attention mechanism 
may learn less meaningful weights 
on sparse recommendation datasets.
We also discussed the advantages and disadvantages of the three experimental settings used in our experiments.
We think \CUS is more suitable for the sequential 
recommendation problem compared to \CUT and \LOS.
However, how to standardize the experimental settings and 
evaluation metrics is still open to future research. 

%
%
%
%

\vspace{-5pt}
\section*{Acknowledgement}
\label{app}

This project was made possible, in part, by support from the National
Science Foundation under Grant Number IIS-1855501 and
IIS-1827472, and from National Library of Medicine under Grant Number 1R01LM012605-01A1. 
Any opinions, findings, and conclusions or recommendations
expressed in this material are those of the authors and do
not necessarily reflect the views of the funding agencies.

\vspace{-10pt}
\bibliographystyle{IEEEtran}
\bibliography{paper}

\vfill
\vspace*{-3\baselineskip}

%
\begin{IEEEbiographynophoto}{Bo~Peng}
 received his M.S. degree from the Department of Computer and Information Science, 
 Indiana University–Purdue University, Indianapolis, in 2019. He is currently a 
 Ph.D. student at the Computer
 Science and Engineering Department, The Ohio State University.
 His research interests include machine learning, data mining and their applications in
 recommender systems and graph mining.

\end{IEEEbiographynophoto}
\vspace*{-2.5\baselineskip}
\begin{IEEEbiographynophoto}{Zhiyun~Ren}
  received her Ph.D. degree from the Department of Computer Science, 
  George Mason University, Fairfax, in 2019. She is currently 
  a Postdoctoral Scholar at the Biomedical Informatics Department, The Ohio State University. 
  Her research interests include machine learning, data mining and their applications 
  in learning analytics, recommender systems and biomedical informatics. 
\end{IEEEbiographynophoto}
\vspace*{-2.5\baselineskip}
\begin{IEEEbiographynophoto}{Srinivasan~Parthasarathy}
received his Ph.D. degree from the Department of Computer Science,
University of Rochester, Rochester, in 1999. He is currently 
a Professor at the Computer Science and Engineering Department, and 
the Biomedical Informatics Department, The Ohio State University.
His research is on high performance data analytics, graph analytics and network science, and machine learning and database systems.

\end{IEEEbiographynophoto}
\vspace*{-2.5\baselineskip}
\begin{IEEEbiographynophoto}{Xia~Ning}
  received her Ph.D. degree from the Department of Computer Science \& Engineering,
  University of Minnesota, Twin Cities, in 2012. She is currently 
  an Associate Professor at the Biomedical Informatics Department, and the Computer
  Science and Engineering Department, The Ohio State University. Her
  research is on data mining, machine learning and artificial intelligence with applications 
  in recommender systems, drug discovery and medical informatics. 
\end{IEEEbiographynophoto}
\vfill

\appendices

\setcounter{table}{0}
\renewcommand{\thetable}{A\arabic{table}}

\section{Parameter Study of \SASRec \LOS }
\label{sec:sasrec:parameter:comics}

\begin{table}[!h]
\footnotesize
  \caption{\mbox{Parameter Study of \SASRec on Comics  in \LOS}}
    \label{tbl:paraSASComics}
  \centering
  \begin{threeparttable}
      \begin{tabular}{
        @{\hspace{0pt}}c@{\hspace{10pt}}
        @{\hspace{10pt}}r@{\hspace{10pt}}
        @{\hspace{10pt}}r@{\hspace{10pt}}
        @{\hspace{10pt}}r@{\hspace{10pt}}
        @{\hspace{10pt}}r@{\hspace{10pt}}
    @{\hspace{10pt}}r@{\hspace{0pt}}
        }
        \toprule
        parameter & $d$ & $n$ & $h$ & Recall@5 & Recall@10 \\
        \midrule
        \multirow{4}{*}{{{\centering $d$}}}
        & 200 & 600 & 1 & 0.2543 & 0.3209\\
        & 400 & 600 & 1 & 0.2611 & 0.3250\\
        & \underline{600} & \underline{600} & \underline{1} 
        & \underline{0.2750} 
        & \underline{0.3353}\\
        & 800 & 600 & 1 & 0.0374 & 0.0588\\
        \midrule
        \multirow{4}{*}{{{\centering $n$}}}
        & 600 & 200 & 1 & 0.2714 & 0.3333\\
        & 600 & 400 & 1 & 0.0459 & 0.0737\\
        & \underline{600} & \underline{600} & \underline{1}
        & \underline{0.2750} 
        & \underline{0.3353}\\
        & 600 & 800 & 1 & OOM & OOM\\
        \midrule
        \multirow{3}{*}{{{\centering $h$}}}
        & \underline{600} & \underline{600} & \underline{1}
        & \underline{0.2750} 
        & \underline{0.3353}\\
        & 600 & 600 & 2 & 0.0357 & 0.0587\\
        & 600 & 600 & 4 & OOM & OOM\\
        \bottomrule
      \end{tabular}
      \begin{tablenotes}[normal,flushleft]
      \begin{scriptsize}
      \item
          In this table, $d$, $n$ and $h$ are
          the dimension of embeddings, 
          maximum sequence length 
          and the number of head of the multi-head attention.
          The best results on validation sets and the corresponding parameters are \underline{underlined}.
          The ``parameter'' column presents the parameters to be studied in each row block.
          The ``OOM'' represents the out-of-memory issue. 
          \par
      \end{scriptsize}
      \end{tablenotes}
  \end{threeparttable}
\end{table}


In this section, we present more details of the parameter study on \SASRec as
discussed in Section~\ref{sec:results:loo} of the main manuscript.
Table~\ref{tbl:paraSASComics} presents the results of \SASRec on the validation set
of Comics in \LOS with respect to different parameters.
In Table~\ref{tbl:paraSASComics}, the ``OOM" represents the out-of-memory issue and
the best results on the validation set and the corresponding parameters
are \underline{underlined}.
Table~\ref{tbl:paraSASComics} shows that the performance of \SASRec changes dramatically
by slightly changing the parameters.
For example, when the dimension of embeddings $d$ is changed from 600 to 800 and all the other parameters are fixed,
the performance of \SASRec on Recall@$10$ decreases dramatically from 0.3353 to 0.0588 (6-fold decrease).
The similar trend could also be found in the other parameters
(i.e., maximum sequence length $n$ and the number of head in the multi-head attention $h$).
For example, when  $h$ is changed from 1 to 2, the performance decreases from 0.3353 to 0.0587
on Recall@$10$ (6-fold decrease).
These results indicate that \SASRec could be very sensitive
to parameters.
Table~\ref{tbl:paraSASComics} also shows that \SASRec requires a large amount of memory
in training, as we got the out-of-memory issue when using large $n$ (i.e., 800) and $h$ (i.e., 4).
This limits the real-application scenarios that \SASRec could be used for.
Our \method methods do not suffer from such memory issues.

\section{Parameters for Reproducibility}
\label{app:sec:para}



In this Section, 
we report the parameters corresponding to the best Recall@$10$ results 
of \methodCorm and all the other baseline methods for 
the sake of reproducibility.
These parameters are identified through tuning on the validation sets. 
Recall that we use the same training and validation sets
in \CUT and \CUS (Fig.~\ref{fig:setting}).
As a result, the best parameters based on the tuning on the validation sets are identical 
in \CUT and \CUS. 

We implement \method in python with pytorch 1.2.0 (https://pytorch.org).
We used Adam optimizer with learning rate 1e-3 and
regularization factor 1e-3 on all the datasets.
The dimension of embeddings $d$,
the number of items in high-order/low-order associations $\LL$/$\M$,
the number of items to calculate recommendation errors during training $\T$,
and the order of item synergies that are specific for each dataset are reported in the 
\methodCorm column of
Table~\ref{tbl:para:ham}.
Our \method implementation is publicly available at~\url{https://github.com/BoPeng112/HAM}.
\begin{table*}[!t]
\footnotesize
  \caption{\mbox{Best Parameters for \methodCorm and Baseline Methods}}
  \centering
  \label{tbl:para:ham}
  \begin{threeparttable}
      \begin{tabular}{
        @{\hspace{10pt}}l@{\hspace{8pt}}
        @{\hspace{8pt}}l@{\hspace{8pt}}
        @{\hspace{8pt}}r@{\hspace{8pt}}
        @{\hspace{8pt}}r@{\hspace{8pt}}
        @{\hspace{8pt}}r@{\hspace{8pt}}
        @{\hspace{8pt}}r@{\hspace{8pt}}
        @{\hspace{8pt}}r@{\hspace{3pt}}
	@{\hspace{8pt}}c@{\hspace{8pt}}
        @{\hspace{3pt}}r@{\hspace{8pt}}
        @{\hspace{8pt}}r@{\hspace{8pt}}
        @{\hspace{8pt}}r@{\hspace{8pt}}
        @{\hspace{8pt}}c@{\hspace{8pt}}
        @{\hspace{3pt}}r@{\hspace{8pt}}
        @{\hspace{8pt}}r@{\hspace{8pt}}
        @{\hspace{8pt}}r@{\hspace{8pt}}
        @{\hspace{8pt}}c@{\hspace{8pt}}
        @{\hspace{3pt}}r@{\hspace{8pt}}
        @{\hspace{8pt}}r@{\hspace{8pt}}
        @{\hspace{8pt}}r@{\hspace{8pt}}
        @{\hspace{8pt}}r@{\hspace{8pt}}
        @{\hspace{8pt}}r@{\hspace{10pt}}
        }
        \toprule
        & \multirow{2}{*}{Dataset} & \multicolumn{5}{c}{\methodCorm} & & \multicolumn{3}{c}{\HGN} 
        & & \multicolumn{3}{c}{\SASRec} & & \multicolumn{5}{c}{\Caser}\\
        \cmidrule(lr){3-7} \cmidrule(lr){9-11} \cmidrule(lr){13-15} \cmidrule(lr){17-21}
        & & $d$ & $\LL$ & $\M$ & $\T$ & $p$ & & $d$ & $L$ & $T$
        & & $d$ & $n$ & $h$
        & & $d$ & $L$ & $T$ & $nv$ & $nh$\\
        \midrule
        \multirow{6}{*}{\rotatebox[origin=c]{90}{{\centering \CUT}}}
                \multirow{6}{*}{\rotatebox[origin=c]{90}{{\centering \CUS}}}
        & CDs      & 400 & 5 & 2 & 3 & 2 & & 200 & 5 & 2 & & 400 & 600 & 1
        & & 200 & 5 & 4 & 2 & 16\\
	& Books    & 400 & 9 & 2 & 7 & 2 & & 400 & 4 & 4 & & 400 & 600 & 1
        & & 200 & 6 & 4 & 2 & 8\\
        & Children & 400 & 6 & 1 & 4 & 3 & & 200 & 2 & 4 & & 400 & 200 & 1
        & & 100 & 4 & 4 & 2 & 16\\
        & Comics   & 400 & 7 & 2 & 5 & 3 & & 200 & 2 & 6 & & 400 & 400 & 1
        & & 100 & 4 & 4 & 2 & 16\\
        & ML-20M   & 400 & 9 & 3 & 2 & 3 & & 100 & 5 & 3 & & 400 & 400 & 4
        & & 100 & 6 & 2 & 4 & 8\\
        & ML-1M    & 400 & 7 & 2 & 3 & 3 & & 100 & 4 & 4 & & 200 & 600 & 1
        & & 200 & 6 & 2 & 2 & 8\\
        \midrule
                \multirow{6}{*}{\rotatebox[origin=c]{90}{{\centering \LOS}}}
        & CDs      & 400 & 4 & 2 & 7 & 2&  & 200 & 4 & 3 & & 400 & 400 & 4
        & & 200 & 4 & 4 & 2 & 16\\
        & Books    & 400 & 9 & 2 & 9 & 2&  & 400 & 2 & 6 & & 400 & 400 & 1
        & & 200 & 5 & 3 & 2 & 8\\
        & Children & 400 & 6 & 1 & 4 & 3&  & 100 & 2 & 5 & & 400 & 200 & 1               
        & & 200 & 4 & 4 & 2 & 8\\
        & Comics   & 400 & 7 & 1 & 5 & 3&  & 200 & 2 & 5 & & 600 & 600 & 1
        & & 200 & 4 & 4 & 2 & 8\\
        & ML-20M   & 400 & 8 & 3 & 3 & 3&  & 100 & 6 & 3 & & 400 & 400 & 4
        & & 200 & 4 & 4 & 2 & 8\\
        & ML-1M    & 400 & 8 & 2 & 2 & 3&  & 100 & 3 & 4 & & 200 & 600 & 2
        & & 200 & 5 & 2 & 2 & 16\\
        \bottomrule
      \end{tabular}
      \begin{tablenotes}[normal,flushleft]
      \begin{scriptsize}
      \item
          In this table, in \methodCorm, $d$, $\LL$/$\M$, $\T$ and $p$ are
          the embedding dimension, number of items in high-order/low-order associations,
          number of items to calculate recommendation errors during training
          and the order of item synergies, respectively.
          In \HGN, $d$, $L$ and $T$ are the embedding dimension, length of the subsequences
          and the number of items used as targets in training.
	  In \SASRec, $d$, $n$ and $h$ are the embedding dimension, maximum sequence length and
          the number of head of the multi-head attention.
          In \Caser, $d$, $L$, $T$, $nv$ and $nh$ are 
          the embedding dimension, the length of the subsequences, the number of items used as targets in training
          the number of vertical filters and the number of horizontal filters, respectively.
          The \methodCorm, \HGN, \SASRec, \Caser columns 
          present the best parameters on validation sets and thus are used 
          in testing for \methodCorm, \HGN, \SASRec and \Caser, respectively.
          \par
      \end{scriptsize}
      \end{tablenotes}
  \end{threeparttable}
\end{table*}

For \HGN, we used the implementation
provided by the authors in github\footnote{\url{https://github.com/allenjack/HGN}}. 
We used the default Adam optimizer with learning rate 1e-3 and
regularization factor 1e-3 on all the datasets.
The dimension of embeddings, denoted as $d$,
the length of the subsequences, denoted as $L$
and the number of items used as targets in training, denoted as $T$, of each dataset
are reported in the \HGN column of Table~\ref{tbl:para:ham}.

%
%

For \SASRec, we used the implementation provided by the authors in
github\footnote{\url{https://github.com/kang205/SASRec}}
as well.
We used the default Adam optimizer with learning rate 1e-3 and
the exponential decay rate for the second-moment estimates beta2 0.98.
The dimension of embeddings, denoted as $d$,
the maximum sequence length, dented as $n$
and the number of heads of the multi-head attention, denoted as $h$, of each dataset
are reported in the \SASRec column of Table~\ref{tbl:para:ham}.

%

For \Caser, we also used the pytorch implementation suggested by the authors in
github\footnote{\url{https://github.com/graytowne/caser_pytorch}}.
We used the default Adam optimizer with learning rate 1e-3
and regularization factor 1e-6.
The dimension of embeddings, denoted as $d$,
the length of the subsequences, denoted as $L$,
the number of items used as targets in training, denoted as $T$,
the number of vertical filters in CNNs, denoted as $nv$, 
and the number of horizontal filters in CNNs, denoted as $nh$, of each dataset
are reported in the \Caser column of Table~\ref{tbl:para:ham}

\end{document}